\documentclass[multphys,vecphys]{svmult}

\usepackage{graphicx}    

\makeindex             

\usepackage{amssymb,amsmath}

\newcommand{\mbf}[1]{{\boldsymbol {#1} }}

\newcommand{\complex}{{\mathbb C}} 
\newcommand{\zed}{{\mathbb Z}} 
\newcommand{\real}{{\mathbb R}} 
\newcommand{\zeds}{{\mathbb Z}} 
\newcommand{\mat}{{\mathbb M}} 
\def\module{{\cal E}}
\newcommand{\id}{{1\!\!1}} 
\def\alg{{\cal A}}

\def\vp{{\mbf p}}
\def\vq{{\mbf q}}
\def\vn{{\mbf n}}
\def\vm{{\mbf m}}

\def\vnu{{\mbf \nu}}
\def\torus{{\mathbf T}}
\def\sphere{{\mathbf S}}

\def\nn{\nonumber}
\newcommand{\tr}[1]{\:{\rm tr}\,#1}
\newcommand{\Tr}[1]{\:{\rm Tr}\,#1}
\def\e{{\,\rm e}\,}

\newcommand{\non}{\nonumber \\}

\def\beq{\begin{equation}}
\def\eeq{\end{equation}}
\def\bea{\begin{eqnarray}}
\def\eea{\end{eqnarray}}
\def\bd{\begin{displaymath}}
\def\ed{\end{displaymath}}

\def\DD{{\rm D}}
\def\dd{{\rm d}}

\def\ii{{\,{\rm i}\,}}

\def\bfp{{\vec p\,}}
\def\bfq{{\vec q\,}}

\begin{document}

\title*{LECTURES ON TWO-DIMENSIONAL\\ NONCOMMUTATIVE GAUGE THEORY\\
2. Quantization}

\titlerunning{Two-Dimensional Noncommutative Gauge Theory}
\author{L.D. Paniak\inst{1}\and
R.J. Szabo\inst{2}}
\institute{Michigan Center for Theoretical Physics, University of
Michigan, Ann Arbor, Michigan 48109-1120, U.S.A.
\texttt{paniak@umich.edu}
\and Department of Mathematics, Heriot-Watt University, Scott
Russell Building, Riccarton, Edinburgh EH14 4AS, U.K.
\texttt{R.J.Szabo@ma.hw.ac.uk}}
%
%
\maketitle

\begin{minipage}{11cm}
\small \baselineskip=11pt

These notes comprise the second part of two articles devoted to the
construction of exact solutions of noncommutative gauge theory in
two spacetime dimensions. Here we shall deal with the quantum field
theory. Topics covered include an investigation of the symmetries of
quantum gauge theory on the noncommutative torus within the path
integral formalism, the derivation of the exact expression for the
vacuum amplitude, and the classification of instanton contributions. A
section dealing with a new, exact combinatorial solution of gauge
theory on a two-dimensional fuzzy torus is also included.

\hfill

Based on invited lectures given by the second author at the {\it 2nd
  Summer School in Modern Mathematical Physics}, September 1--12 2002,
  Kopaonik, Yugoslavia; at the {\it European IHP Network Conference on Quantum
  Gravity and Random Geometry}, September~7--15 2002, Orthodox
  Academy of Crete, Kolympari, Greece; and at the {\it International
  Workshop on Quantum Field Theory and Noncommutative Geometry},
  November~26--30 2002, Tohoku University, Sendai, Japan. To be
  published in the Workshop proceedings by Springer-Verlag as Lecture
  Notes in Physics.

\end{minipage}

\hfill

\hfill

\centerline{\small\sf MCTP--03--20 , HWM--03--7 , EMPG--03--08 ,
 hep-th/0304268 ; April 2003}

\setcounter{equation}{0}
\section{Introduction}
\label{sec:1}

These lecture notes continue the study of noncommutative gauge theory
in two dimensions which was begun in~\cite{ncproc1} at the classical
level. In this second part we shall deal with matters concerning the
quantization of these gauge theories, and in particular demonstrate
how to explicitly obtain non-perturbative solutions. Some background
and motivation for dealing with this particular class of models may be
found in~\cite{ncproc1} and won't be repeated here. Various aspects of
two-dimensional noncommutative gauge theory have been studied over the
past few years in~\cite{poly}--\cite{inprep2}. In the present article we
shall only analyze the vacuum amplitudes of these theories. More
general gauge invariant correlation functions are studied
in~\cite{bnt,gsv,Bietenholz:2002ch},\cite{bnt2}--\cite{inprep2}.
Reviews on noncommutative field theory pertinent to the present
material may be found in~\cite{ks}--\cite{sz1}. A detailed review of
ordinary Yang-Mills theory in two dimensions is given
in~\cite{cmr}. All relevant mathematical details and properties of the
classical noncommutative gauge theory may be found in~\cite{ncproc1}
and are briefly reviewed in section~\ref{subsec:1.2} below.

\subsection{How to Solve Yang-Mills Theory in Two Dimensions}
\label{subsec:1.1}

When one comes to the issue of quantizing noncommutative gauge
theory in two dimensions, one is naively faced with a plethora of
possibilities. The commutative version of this theory has a long
history as an exactly solvable quantum field theory, and as such is
explicitly solvable by many different techniques. We will therefore
begin with a brief run through of the various methods that may be
used to solve ordinary Yang-Mills theory, and elucidate on the
possibilities of extending them to the noncommutative setting.

\paragraph{Heat Kernel/Group Theory Methods}

One of the most profound features of two-dimensional Yang-Mills theory
is the interplay between the two-dimensional geometry on which it is
defined and the representation theory of its structure group~\cite{cmr}. As
will be reviewed in section~\ref{subsec:3.1}, the propagator between two
states may be easily written down in terms of the standard heat kernel
on the group manifold of the structure group, and from this the vacuum
amplitude and Wilson loops on arbitrary geometries may be
extracted. However, these techniques are not readily available in
the noncommutative case for several reasons. First and foremost is the
lack of a notion of structure group in the noncommutative
setting. While there is a well-defined gauge group, it mixes spacetime
and internal colour symmetries through noncommutative gauge
transformations and there is no clear separation of spacetime and
internal degrees of freedom. Secondly, a Hamiltonian formalism is not
available because making time a noncommutative coordinate causes
problems with unitarity and the overall interpretation of
time-evolution in these systems. While this approach in the
commutative case will play a crucial role in the foregoing line of
development, it is not the one that will be {\it a priori} used
analyse the quantum field theory. The group theory approach in the
noncommutative setting has been analysed recently in~\cite{inprep2}.

\paragraph{Integrability}

The fact that Yang-Mills theory is exactly solvable in two-dimensions
is intimately connected with the fact that it is related to an
integrable system~\cite{YMintegrable}. It is possible to relate dynamics in
this theory to that of certain one-dimensional gauged matrix models which are
related to Calogero-Moser systems~\cite{MinPol,GPSS}. While the integrability
of the noncommutative counterpart may be established to a certain
extent~\cite{ncproc1}, it is not clear what integrable structure
underlies this system. This line of attack therefore does not
immediately lead to an appropriate generalization.

\paragraph{Semi-Classical Methods}

One way to understand the exact solvability of the two-dimensional
gauge theory is through the observation that its partition function
and observables are given exactly by their semi-classical
approximation~\cite{Wittensemi}. This is related to the fact that ordinary
Yang-Mills theory can be recast as a cohomological field theory in two
dimensions. These properties {\it do} generalize to the noncommutative
setting with some care, as we discuss in section~\ref{sec:2}. In fact,
these techniques will be the focal point of much of this article. They
have been recently applied in~\cite{inprep2} to explicitly
compute the correlation functions of open Wilson line operators.

\paragraph{Lattice Regularization}

Discretizing spacetime also provides a fruitful way of tackling the
problem and is at the very heart of the group theory methods mentioned
above~\cite{Migdal,Rusakov}. While a lattice formulation of noncommutative
gauge theory is available~\cite{amns}, it is much more complicated than its
commutative version because the nice self-similarity property possessed by the
latter is ruined by the inherent non-locality of the former. Nonetheless, we
have succeeded in explicitly solving the lattice model in two
dimensions at finite cutoff, and this is new material which will be
presented in detail in section~\ref{sec:5}. We shall therefore
postpone further discussion of this approach until then.

\paragraph{Relations to Other Field Theories}

Besides its relationship with a cohomological gauge theory,
two-dimensional quantum Yang-Mills theory may also be related to
various other field theories in certain limits, such as
three-dimensional Chern-Simons theory and two dimensional conformal
field theory~\cite{WittenQ}. These connections can be used to give explicit
formulas for the volumes of the moduli spaces of representations of fundamental
groups of two-dimensional surfaces. As we discuss in
section~\ref{sec:4}, some of these volumes are also effectively
computable in the noncommutative setting. These ideas can also all be
cast into the formalism of abelianization~\cite{GPSS}, a technique that relies
heavily upon the presence of a well-defined structure group. However, it is not
clear what sort of mathematical structures one should find in general
and these further connections remain an interesting, as yet unexplored
area of this subject.

\subsection{Background from Part 1}
\label{subsec:1.2}

As we have mentioned above, the solution of the quantum gauge theory
will be determined in large part by the very structure of the
classical solutions of the field theory. This is described at length
in~\cite{ncproc1}. To keep the presentation of the present article
reasonably self-contained and to set some notation, we shall briefly
summarize the classical solutions of gauge theory on a noncommutative
torus in two-dimensions that were obtained in~\cite{ncproc1}. The
classical action is defined on a fixed Heisenberg module
$\module_{p,q}$ over the algebra $\alg_\theta$ of functions on the
noncommutative torus of fixed topological numbers $(p,q)\in\zed^2$, with
$q$ the Chern number, $\dim{\cal E}_{p,q}=p-q\,\theta>0$, and $N={\rm
  gcd}(p,q)$ the rank of the gauge theory. It is given explicitly by
\beq
S[A]=\frac1{2g^2}\,\Tr\left[\nabla_1\,,\,\nabla_2\right]^2
=\frac1{2g^2}\,\int\dd^2x~\tr^{~}_N\left(F_A(x)-\frac{2\pi\,q}
{p-q\,\theta}\right)^2 \ ,
\label{YMactiondef}\end{equation}
where $\nabla=\partial+A$ is a connection on $\module_{p,q}$ and $F_A$
is the corresponding field strength. In the first equality of
(\ref{YMactiondef}), $\Tr$ is the canonical trace on the endomorphism
algebra ${\rm End}(\module_{p,q})$. In the second equality the
integration extends over the two-dimensional, unit area square torus
$\torus^2$, $\tr^{~}_N$ is the usual $N\times N$ matrix trace, and the
constant subtraction corresponds to the constant curvature of the
module $\module_{p,q}$.

Classical solutions of this gauge theory are in a one-to-one
correspondence with the direct sum decompositions
\beq
\module_{p,q}={\bigoplus}_k\,\module_{p_k,q_k}
\label{directsumdecomp}\end{equation}
of the given Heisenberg module into projective submodules. These are
characterized by {\it partitions} $(\bfp,\bfq)=\{(p_k,q_k)\}$ of the
topological numbers $(p,q)$ satisfying the constraints
\bea
p_k-q_k\theta&>&0 \ , \non{\sum}_k\,\left(p_k-q_k\theta\right)&=&
p-q\,\theta \ , \non{\sum}_k\,q_k&=&q \ .
\label{partitiondef}\eea
In addition, to avoid overcounting, it is sometimes useful to
impose a further ordering constraint $p_k-q_k\theta\leq
p_{k+1}-q_{k+1}\theta~~\forall k$, and regard any two partitions
as the same if they coincide after rearranging their components
according to this ordering. We may then characterize the
components of a partition by integers $\nu_a>0$ which are defined as the
number of partition components that have the $a^{\rm th}$ least
dimension $p_a-q_a\theta$. The integer
\beq
|\vnu|={\sum}_a\,\nu_a
\label{vnudef}\end{equation}
is then the total number of components in the given partition. The
noncommutative Yang-Mills action (\ref{YMactiondef}) evaluated on a
classical solution, with corresponding partition $(\vp,\vq)$, is given by
\beq
S(\vp,\vq)=\frac{2\pi^2}{g^2}\,{\sum}_k\,\left(p_k-q_k\theta\right)
\left(\frac{q_k}{p_k-q_k\theta}-\frac q{p-q\,\theta}\right)^2 \ .
\label{NCYMpartition}\eeq

\subsection{Outline}
\label{subsec:1.3}

The outline of material in the remainder of this paper is as
follows. In section~\ref{sec:2} we will carefully define the quantum
theory, examine a deep ``hidden'' supersymmetry of it, and prove that
the partition function and observables are all given exactly by their
semi-classical approximation. In section~\ref{sec:3} we will derive
an exact, analytical expression for the partition function of gauge
theory on the noncommutative torus in two dimensions, and use it to
analyse precisely how noncommutativity alters the properties of
Yang-Mills theory on $\torus^2$. In
section~\ref{sec:4} we will describe how to organize the
non-perturbative expression for the vacuum amplitude into a sum over
contributions from (unstable) instantons of the two-dimensional gauge
theory, and compare with analogous expressions obtained on the
noncommutative plane. This paves the way for our analysis in
section~\ref{sec:5} which deals with the matrix model/lattice
formulation of noncommutative gauge theory in two dimensions. We will
present here a new, exact expression for the partition function on the
fuzzy torus, and describe the scaling limits which map this model
onto the continuum gauge theory.

\section{Quantum Gauge Theory on the Noncommutative Torus}
\label{sec:2}

In this section we will carefully define the quantum gauge theory
within the path integral formalism. We will show that it admits a
natural interpretation as a phase space path integral of an
infinite-dimensional Hamiltonian system, the noncommutative Yang-Mills
system. In this formulation a particular cohomological symmetry of the
quantum theory is manifest, which leads immediately to the property
that the partition function is given exactly by its semi-classical
approximation. While naively the vacuum amplitude may seem to merely
produce uninteresting determinants, the non-trivial topology of the
torus provides a rich analytic structure (through large gauge
transformations). For some time we will neglect the constant curvature
subtraction in the action (\ref{YMactiondef}), and simply reinstate it when we
come to the derivation of the exact formula for the partition function. This is
possible to do because of the Morita invariance of the gauge
theory~\cite{ncproc1}.

\subsection{Definition}
\label{subsec:2.1}

The quantum field theory is defined formally through the functional
integral
\beq
Z=\int\DD A~\e^{-S[A]} \ ,
\label{NCYMpartitiondef}\eeq
where $S[A]$ is the noncommutative Yang-Mills action
(\ref{YMactiondef}). The integration in (\ref{NCYMpartitiondef}) is over
the space ${\cal C}={\cal C}(\module)$ of compatible connections on a
given fixed Heisenberg module $\module=\module_{p,q}$. Since the
action is gauge invariant, the integration measure $\DD A$ must be carefully
defined so as to select only gauge orbits of the field
configurations. There is a very natural way to define this measure in
the present situation. As discussed in~\cite{ncproc1}, the
noncommutative Yang-Mills system naturally defines a Hamiltonian
system with moment map $\mu[A]=F_A$, so that the Yang-Mills action is
the square of the moment map, $S[A]=\Tr\mu[A]^2/2g^2$. The
gauge-invariant symplectic form
\beq
\omega[\alpha,\beta]=\Tr\alpha\wedge\beta \ , ~~
\alpha,\beta\in\Omega^1(\module) \ ,
\label{omegaab}\eeq
is defined on the tangent space to $\cal C$, which is identified as
the space $\Omega^1({\cal E})={\rm
  End}(\module)\otimes\bigwedge^1{\cal L}^*$ with $\cal L$ the
(centrally extended) Lie algebra of the translation group acting on
$\torus^2$. We have defined
$\alpha\wedge\beta\equiv\alpha_1\beta_2-\alpha_2\beta_1$ with respect
to an orthonormal basis of $\cal L$.

We now let
\beq
\dd A=\prod_{a,b=1}^N\,\prod_{x\in\torus^2}\dd A_1^{ab}(x)~
\dd A_2^{ab}(x)
\label{Feynmeasure}\eeq
be the usual, formal (gauge non-invariant) Feynman path integral
measure on $\cal C$, and let $\psi$ be
the odd generators of the infinite-dimensional superspace ${\cal
  C}\oplus\Pi\Omega^1(\module)$ with corresponding functional Berezin
measure $\dd A~\dd\psi$, where $\Pi$ is the parity reversion
operator. We may then define
\beq
\DD A=\dd A~\int\dd\psi~\e^{-\ii\omega[\psi,\psi]} \ ,
\label{DDAsympldef}\eeq
where here and in the following we will absorb the infinite volume of
the group ${\cal G}={\cal G}(\module)$ of gauge transformations on
$\cal C$ (determined by the Haar measure $\dd\nu$ on $\cal G$ induced
by the inner product $(\lambda,\lambda')=\Tr\lambda\lambda'$,
$\lambda,\lambda'\in{\rm End}(\module)$), by which (\ref{DDAsympldef})
should be divided. By construction, this measure is gauge-invariant and
coincides with the functional Liouville measure associated to the
infinite-dimensional dynamical system. An infinitesimal gauge transformation
$A\mapsto A+[\nabla,\lambda]$, $\lambda\in{\rm End}(\module)$ on $\cal
C$ naturally induces the transformation $\psi\mapsto\psi+[\lambda,\psi]$ on
its tangent space, under which (\ref{omegaab}) is invariant. In this
setting the partition function (\ref{NCYMpartitiondef}) is naturally
defined as a phase space path integral. Note that, since the fermion
fields $\psi$ appear only quadratically in (\ref{DDAsympldef}), this
measure coincides with that of its commutative counterpart at $\theta=0$.

While this definition is very natural in the present context, we
should demonstrate explicitly that it coincides with the more
conventional gauge field measure obtained from the standard
Faddeev-Popov gauge fixing procedure. The basic point is that the
measure (\ref{DDAsympldef}) has the following requisite property. Let
$\pi:{\cal C}\to{\cal C}/{\cal G}$ be the projection onto the quotient space
of $\cal C$ by the gauge group $\cal G$. Then the quotient measure
$\DD A$ on ${\cal C}/{\cal G}$ is the measure which satisfies $\dd
A=\pi^*(\DD A)~\dd\nu$. The Faddeev-Popov procedure constructs
$\DD A$ by introducing the standard fermionic ghost field
$c\in\Pi\Omega^0(\module)=\Pi\,{\rm End}(\module)$, and the anti-ghost
multiplet consisting of a fermionic field
$\overline{c}\in\Pi\Omega^0(\module)$ and a bosonic field
$w\in\Omega^0(\module)$, along with the BRST transformation laws
\bea
\delta A&=&-[\nabla,c] \ , \non\delta c&=&\mbox{$\frac12$}\,[c,c]
\ , \non\delta\overline{c}&=&\ii w \ , \non\delta w&=&0
\label{BRSTtransf}\eea
obeying $\delta^2=0$.

The gauge-fixing term is given by $I=-\delta V$ for a suitable
functional $V$ of the BRST field multiplet. For this, we write a
generic connection $\nabla$ in the neighbourhood of a representative
$\nabla^0=\partial+A^0\in{\cal C}$ of its gauge orbit as $\nabla=\nabla^0+B$,
and make the local choice $V=-\Tr\,\overline{c}\,\nabla^0\cdot
B$, where we have defined
$\alpha\cdot\beta\equiv\alpha_1\beta_1+\alpha_2\beta_2$ for cotangent
vectors $\alpha,\beta\in\Omega^1(\module)$. This produces
\beq
I=\Tr\left(\ii w\,\nabla^0\cdot B-\overline{c}\,\nabla^0\cdot
\nabla c\right) \ ,
\label{Iterm}\eeq
and the gauge fixed path integral measure is then defined by
\beq
\DD A=\dd A^0~\int\dd B~\dd c~\dd\overline{c}~\dd w~\e^{-I} \ .
\label{gaugefixmeas}\eeq
Formally integrating over the bosonic field $w$ and the Grassmann
fields $c,\overline{c}$ gives
\beq
\DD A=\dd A^0~\int\dd B~\delta\left(\nabla^0\cdot B\right)\,
\det\nabla^0\cdot\nabla \ .
\label{wccint}\eeq
The integration over $B$ enforces the gauge condition $\nabla^0\cdot B=0$
on the quantum field theory with the choice (\ref{Iterm}) of gauge-fixing
term. Since $\delta(\nabla^0\cdot
B)=\delta(B)/|\det\nabla^0\cdot\nabla^0|$, the resulting ratio of
determinants after integrating out $B$ in (\ref{wccint}) coincides
exactly with the determinant induced by integrating out $\psi$ in
(\ref{DDAsympldef}). Thus the elementary measure defined by the
symplectic structure of $\cal C$ coincides with that of the usual
Faddeev-Popov gauge-fixing procedure.

\subsection{The Cohomological Gauge Theory}
\label{subsec:2.2}

We will now describe a remarkable cohomological symmetry of the
partition function (\ref{NCYMpartitiondef}), with path integration measure
(\ref{DDAsympldef}), which will be the crux of much of our ensuing
analysis of the quantum gauge theory. For this, we linearize the Yang-Mills
action in the field strength $F_A$ via a functional Gaussian integral
transformation defined by an auxilliary field $\phi\in\Omega^0(\module)$ as
\beq
Z=\int\dd\phi~\e^{-\frac{g^2}2\,\Tr\phi^2}~\int\dd A~\dd\psi~
\e^{-\ii\Tr(\psi\wedge\psi-\phi\,F_A)} \ .
\label{auxphiint}\eeq
Note that because of the quadratic form of the action in
(\ref{auxphiint}), the only place where noncommutativity is buried is
in $F_A$. This is one of the features that makes this quantum field
theory effectively solvable.

The basic field multiplet $(A,\psi,\phi)$ possesses a ``hidden
supersymmetry'' that resides in the cohomology of the operator
\beq
{\sf Q}_\phi=\Tr\left(\psi\cdot\mbox{$\frac\delta{\delta A}$}+
[\nabla,\phi]\cdot\mbox{$\frac\delta{\delta\psi}$}\right)
\label{Qphidef}\eeq
which generates the transformations
\bea
[{\sf Q}_\phi,A]&=&\psi \ , \non\{{\sf Q}_\phi,\psi\}&=&
[\nabla,\phi] \ , \non {}[{\sf Q}_\phi,\phi]&=&0 \ .
\label{Qphitransf}\eea
The crucial property of the operator ${\sf Q}_\phi$ is that it is
nilpotent precisely on gauge invariant field configurations,
\beq
({\sf Q}_\phi)^2=\delta_\phi \ ,
\label{Qphinilpotent}\eeq
where $\delta_\lambda A=[\nabla,\lambda]$ is an infinitesimal gauge
transformation with gauge parameter $\lambda\in{\rm
  End}(\module)$. Furthermore, the linearized action on ${\cal
  C}\oplus\Pi\Omega^1(\module)$, for fixed $\phi$, is closed under
${\sf Q}_\phi$,
\beq
{\sf Q}_\phi\Tr(\psi\wedge\psi-\phi\,F_A)=0 \ ,
\label{Qphiclosed}\eeq
which follows from the Hamiltonian flow equations for the moment map
$\mu[A]=F_A$ and symplectic structure
(\ref{omegaab})~\cite{ncproc1}. Thus the partition function
(\ref{auxphiint}) defines a cohomological gauge theory with
supersymmetry (\ref{Qphitransf}) and $(A,\psi,\phi)$ the basic field
multiplet of topological Yang-Mills theory in two dimensions~\cite{Wittensemi}.

The nilpotency property (\ref{Qphinilpotent}) implies that the operator ${\sf
  Q}_\phi$ is simply the BRST supercharge, acting in the quantum field
  theory (\ref{auxphiint}), which generates the
  transformations (\ref{BRSTtransf}). Gauge fixing in this setting
  amounts to introducing additional anti-ghost multiplets analogous to
  those that were used in the previous subsection for BRST
  quantization. We shall return to this point in
  section~\ref{subsec:4.1}. From a more formal perspective, ${\sf
  Q}_\phi$ is the Cartan model differential for the $\cal
  G$-equivariant cohomology of $\cal C$~\cite{SzBook}. The second term in the
action of (\ref{auxphiint}) is the $\cal G$-equivariant extension of the
  moment map on $\cal C$, the integration over $A,\psi$ defines an
  equivariant differential form in $\Omega_{\cal G}({\cal C})$, and
  the integral over $\phi$ defines equivariant integration of such
  forms. In this way, as we explain in the next subsection, the
  cohomological symmetry of the quantum field theory will lead to a
  localization theorem for the partition function. Fundamentally, the
  localization points correspond to the BRST fixed points of the
  anti-ghost multiplets.

\subsection{Localization of the Partition Function}
\label{subsec:2.3}

We now come to the fundamental consequence of the hidden supersymmetry
of the previous subsection. Let $\alpha$ be any gauge invariant
functional of the fields of (\ref{auxphiint}), i.e. $({\sf
  Q}_\phi)^2\alpha=0$, and consider the one-parameter family of
partition functions defined by
\beq
Z_t=\int\dd\phi~\e^{-\frac{g^2}2\,\Tr\phi^2}~\int\dd A~\dd\psi~
\e^{-\ii\Tr(\psi\wedge\psi-\phi\,F_A)-t\,{\sf Q}_\phi\alpha}
\label{partfntdef}\eeq
with $t\in\real$. The $t\to0$ limit of (\ref{partfntdef}) is just the
partition function (\ref{auxphiint}) of interest, $Z=Z_0$. The
remarkable feature of (\ref{partfntdef}) is that it is independent of
the parameter $t\in\real$. This follows from the Leibnitz rule for the
functional derivative operator ${\sf Q}_\phi$, the supersymmetry
(\ref{Qphiclosed}) of the (equivariantly extended) action, and the
gauge invariance of $\alpha$ which, along with a formal functional
integration by parts over the superspace ${\cal
  C}\oplus\Pi\Omega^1(\module)$, can be easily used to show that $\partial
Z_t/\partial t=0$. This is a basic cohomological property of the
noncommutative quantum field theory. Adding a supersymmetric ${\sf
  Q}_\phi$-exact term to the action deforms it without changing the
value of the functional integral. It follows that the path integral
(\ref{auxphiint}) can be alternatively evaluated as the $t\to\infty$ limit of
the expression (\ref{partfntdef}). It thereby receives contributions
from only those field configurations which obey the equations ${\sf
  Q}_\phi\alpha=0$.

At this stage we need to specify an explicit form for
$\alpha$. Different choices will localize the partition function onto
different components in field space, but the final results are (at
least superficially) all formally identical. A convenient choice is
$\alpha=\Tr\psi\cdot[\nabla,F_A]$, for which the $t\to\infty$ limit of
(\ref{partfntdef}) yields
\bea
Z&=&\int\dd A~\dd\psi~\e^{-\Tr\big(\ii\psi\wedge\psi+\frac1{2g^2}\,(F_A)^2
\big)}~\lim_{t\to\infty}\,\e^{-\frac{t^2}{2g^2}\,\Tr\big[\nabla\,
\stackrel{\scriptstyle\cdot}{\scriptstyle ,}\,
[\nabla,F_A]\,\big]^2}\non&&\times\,({\rm fermions})
\label{Ztinfty}\eea
after performing the functional Gaussian integration over
$\phi$. These arguments of course assume formally that the original
action has no flat directions, but in the present case this is not a
problem since it has a nondegenerate kinetic energy. The
additional terms involving the Grassmann fields $\psi$ in
(\ref{Ztinfty}) formally yield a polynomial function in the parameter
$t$ after integration, and their precise form is not important. What
is important here is the quadratic term in $t$, which in the limit
implies that the functional integral vanishes everywhere except near
those points in $\cal C$ which are solutions of the equations
\beq
\big[\nabla\stackrel{\cdot}{,}[\nabla,F_A]\,\big]=0 \ ,
\label{loceq}\eeq
where we have used positivity of the trace $\Tr$ on ${\rm
  End}(\module)$. By using the Leibnitz rule and the integration by
  parts property $\Tr[\nabla,\lambda]=0$ this equation implies
\beq
0=\Tr F_A\big[\nabla\stackrel{\cdot}{,}
[\nabla,F_A]\,\big]^2=-\Tr[\nabla,F_A]\cdot[\nabla,F_A] \ .
\label{loceqLeibnitz}\eeq
By using non-degeneracy of the trace on ${\rm End}(\module)$ we arrive
finally at
\beq
[\nabla,F_A]=0 \ ,
\label{NCYMeom}\eeq
which are just the classical equations of motion of the original
noncommutative gauge theory.

We have thereby formally shown that the partition function of
noncommutative gauge theory in two dimensions receives
contributions {\it only} from the space of solutions of the
noncommutative Yang-Mills equations. As we reviewed in
section~\ref{subsec:1.2}, each such solution corresponds to a
partition $(\vp,\vq)$, obeying the constraints (\ref{partitiondef}),
of the topological numbers $(p,q)$ of the given Heisenberg module
$\module=\module_{p,q}$ on which the gauge theory is
defined. Symbolically, the partition function is therefore given by
\beq
Z=Z_{p,q}=\sum_{\stackrel{\scriptstyle{\rm partitions}}
{\scriptstyle(\vp,\vq)}}W(\vp,\vq)~\e^{-S(\vp,\vq)} \ .
\label{Zpqlocclass}\eeq
This result expresses the fact that quantum noncommutative gauge
theory in two dimensions is given exactly by a sum over contributions
from neighbourhoods of the stationary points of the Yang-Mills action
(\ref{YMactiondef}). The Boltzmann weight $\e^{-S(\vp,\vq)}$
involving the action (\ref{NCYMpartition}) gives the contribution to
the path integral (\ref{NCYMpartitiondef}) from a classical solution,
while $W(\vp,\vq)$ encode the quantum fluctuations about each
stationary point. These latter terms may in principle be determined
from (\ref{Ztinfty}) by carefully integrating out the fermion fields and
evaluating the functional fluctuation determinants that
arise. However, these determinants are not effectively computable and
are rather cumbersome to deal with. In the next section our main goal
will be to devise an alternative method to extract these quantum
fluctuation terms and hence the exact solution of the noncommutative
quantum field theory.

\section{Exact Solution}
\label{sec:3}

In this section we will present the exact solution of gauge theory on
the noncommutative torus in two dimensions. We will start by recalling
some well-known facts about ordinary Yang-Mills gauge theory in two
dimensions, and show how it can be cast precisely into the form
(\ref{Zpqlocclass}). From this we will then extract the exact
expression in the general noncommutative case. Our techniques will
rely heavily on the full machinery of the geometry of the noncommutative torus.

\subsection{The Torus Amplitude}
\label{subsec:3.1}

\begin{figure}
\centering
\includegraphics[height=2cm]{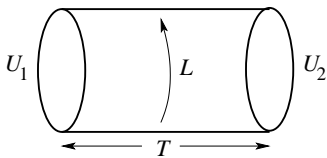}
\caption{Quantization of Yang-Mills theory on a spatial circle of
  circumference $L$ yields the propagation amplitude between two
  states characterized by holonomies $U_1$ and $U_2$ in time $T$.}
\label{cylinder}\end{figure}

The vacuum amplitude for ordinary Yang-Mills theory on the torus
$\torus^2$ with structure group $U(p)$ and generators $T^a$,
$a=1,\dots,p^2$ may be obtained as follows~\cite{cmr}. Let
us consider the physical Hilbert space, in canonical quantization, for
gauge theory on a cylinder $\real\times\sphere^1$~(fig.~\ref{cylinder}). In two
dimensions, Gauss' law implies that the physical state wavefunctionals
$\Psi_{\rm phys}[A]=\Psi[U]$ depend only on the holonomy
$U={\rm P}\,\exp\ii\int_0^L\dd x~A_1(x)$ of the gauge connection
around the cycle of the cylinder. By gauge invariance, $\Psi$ furthermore
depends only on the conjugacy class of $U$. It follows that the
Hilbert space of physical states is the space of $L^2$-class
functions, invariant under conjugation, with respect to the invariant Haar
measure $[\dd U]$ on the unitary group $U(p)$,
\beq
{\cal H}_{\rm phys}=L^2\big(U(p)\big)^{{\rm Ad}\big(U(p)\big)} \ .
\label{Hilbertphys}\eeq
By the Peter-Weyl theorem, it may be decomposed into the unitary
irreducible representations $R$ of $U(p)$ as ${\cal H}_{\rm
  phys}\cong\bigoplus_RR\otimes\overline{R}$. The representation basis
of this Hilbert space is thereby provided by characters in the
unitary representations, such that the states $|R\rangle$ have
wavefunctions
\beq
\langle U|R\rangle=\chi^{~}_R(U)=\tr^{~}_R\,U \ .
\label{chiRUdef}\eeq

The Hamiltonian acting on the physical state wavefunctions $\Psi[U]$
is given by the Laplacian on the group manifold of $U(p)$,
\beq
H=\mbox{$\frac{g^2}2$}\,L\,\tr\left(\mbox{$U\,\frac\partial
{\partial U}$}\right)^2 \ ,
\label{Hamphys}\eeq
and it is thereby diagonalized in the representation basis as
\beq
H\chi^{~}_R(U)=\mbox{$\frac{g^2}2$}\,L\,C_2(R)~\chi^{~}_R(U) \ ,
\label{Hamdiag}\eeq
where $C_2(R)$ is the eigenvalue of the quadratic Casimir operator
$C_2=\sum_aT^a\,T^a$ in the representation $R$. From these facts it is
straightforward to write down the cylinder amplitude corresponding to
propagation of the system between two states with holonomies $U_1$ and
$U_2$ in the form (fig.~\ref{cylinder})
\beq
Z_p(T;U_1,U_2)=\langle U_1|\e^{-T\,H}|U_2\rangle=
{\sum}_R\,\chi^{~}_R(U_1)\,\chi^{~}_R(U^\dag_2)~\e^{-\frac{g^2}2
\,LT\,C_2(R)} \ .
\label{cylampl}\eeq
This is just the standard heat kernel on the $U(p)$ group. In keeping
with our previous normalizations, we shall set the area of the
cylinder to unity, $LT=1$. To extract from (\ref{cylampl}) the
partition function of $U(p)$ Yang-Mills theory on the torus, we glue
the two ends of the cylinder together by setting $U_1=U_2=U$ and
integrate over all $U$ by using the fusion rule for the $U(p)$
characters,
\beq
\int[\dd U]~\chi^{~}_{R_1}(VU)\,\chi^{~}_{R_2}(U^\dag W)=\delta_{R_1,R_2}
\,\frac{\chi^{~}_{R_1}(VW)}{\dim R_1} \ ,
\label{fusionrule}\eeq
where $\dim R=\chi^{~}_R(\id)$. This yields the torus vacuum amplitude
\beq
Z_p=\int[\dd U]~Z_p(T;U,U)={\sum}_R\,\e^{-\frac{g^2}2\,C_2(R)} \ .
\label{torusamplgroup}\eeq

We can make the sum over the irreducible unitary representations $R$ of $U(p)$
in (\ref{torusamplgroup}) explicit by using the fact that each $R$ is
labelled by a decreasing set $\vec{n}=(n_1,\dots,n_p)$ of $p$ integers
\beq
+\infty>n_1>n_2>\cdots>n_p>-\infty
\label{ndecrease}\eeq
which are shifted highest weights parametrizing the lengths of the
rows of the corresponding Young tableaux. Up to an irrelevant
constant, the quadratic Casimir can be written in terms of these
integers as
\beq
C_2(R)=C_2(\vec{n})=\sum_{a=1}^p\left(n_a-\frac{p-1}2\right)^2 \ .
\label{C2Rn}\eeq
Since (\ref{C2Rn}) is symmetric under permutations of the $n_a$'s, it
follows that the ordering restriction (\ref{ndecrease}) can be removed
in the partition function (\ref{torusamplgroup}) to write it as a sum
over non-coincident integers as (always up to inconsequential constants)
\beq
Z_p=\sum_{n_1\neq\cdots\neq n_p}\e^{-\frac{g^2}2\,C_2(\vec{n})} \ .
\label{orderinggone}\eeq
We may extend the sums in (\ref{orderinggone}) over {\it all} $\vec
n\in\zed^p$ by inserting the products of delta-functions
\beq
\det_{1\leq a,b\leq p}\,\left(\delta_{n_a,n_b}\right)=
\sum_{\sigma\in S_p}(-1)^{|\sigma|}\,\prod_{a=1}^p
\delta_{n_a,n_{\sigma(a)}} \ .
\label{detdelta}\eeq
The vanishing of the determinant for coincident rows prevents any two
$n_a$'s from coinciding when inserted into the sum.

Because of the permutation symmetry of (\ref{C2Rn}), when inserted
into the partition function (\ref{orderinggone}) the sum in
(\ref{detdelta}) truncates to a sum over conjugacy classes
$[1^{\nu_1}\,2^{\nu_2}\cdots p^{\nu_p}]$ of the symmetric group
$S_p$. They are labelled by {\it partitions} of the rank $p$ of the
gauge theory,
\beq
\nu_1+2\nu_2+\cdots+p\nu_p=p \ ,
\label{nuapartition}\eeq
where $\nu_a$ is the number of elementary cycles of length $a$ in
$[1^{\nu_1}\,2^{\nu_2}\cdots p^{\nu_p}]$. The sign of such a conjugacy
class is $(-1)^{p+|\vec{\nu}|}$ and its order is
$p!/\prod_aa^{\nu_a}\,\nu_a!$, where
$|\vec{\nu}|=\nu_1+\nu_2+\cdots+\nu_p$ is the total number of cycles in
the class. By using this, along with the Poisson resummation formula
\beq
\sum_{n=-\infty}^\infty\e^{-\pi\,g\,n^2-2\pi\ii b\,n}=
\frac1{\sqrt g}~\sum_{q=-\infty}^\infty\e^{-\pi(q-b)^2/g} \ ,
\label{Poissonresum}\eeq
we may bring the vacuum amplitude (\ref{orderinggone}) after some work
into the form~\cite{inprep}
\bea
Z_p&=&\sum_{\vec{\nu}\,:\,\sum_aa\nu_a=p}~\sum_{q_1,\dots,q_{|\vec{\nu}|}=
-\infty}^\infty\e^{\ii\pi\big(|\vec{\nu}|+(p-1)q\big)}\non&&\times\,
\prod_{a=1}^p\frac{\left(g^2a^3/2\pi^2\right)^{-\nu_a/2}}{\nu_a!}~
\e^{-S(\vec{\nu},\vq)} \ ,
\label{Zptotalinst}\eea
where $q=q_1+q_2+\cdots+q_{|\vec{\nu}|}$ and
\bea
S(\vec{\nu},\vq)&=&\frac{2\pi^2}{g^2}\,\left(\,\sum_{k_1=1}^{\nu_1}
\frac{q_{k_1}^2}1+\sum_{k_2=\nu_1+1}^{\nu_1+\nu_2}\frac{q_{k_2}^2}2
+\sum_{k_3=\nu_1+\nu_2+1}^{\nu_1+\nu_2+\nu_3}\frac{q_{k_3}^2}3
\right.\non&&+\left.\cdots+\sum_{k_p=\nu_1+\cdots+\nu_{p-1}+1}^{|\vec{\nu}|}
\frac{q_{k_p}^2}p\right) \ .
\label{actionnuq}\eea

The important feature of the final expression (\ref{Zptotalinst}) is
that it agrees with the expected sum (\ref{Zpqlocclass}) over
classical solutions of the commutative gauge theory on $\torus^2$. For
this, we note that the K-theory group of the ordinary torus is
${\rm K}_0(C(\torus^2))=\zed\oplus\zed$, so that any projective module
$\module=\module_{p,q}$ over the algebra $C(\torus^2)$ of functions on
the torus is determined by a pair of integers
$(p,q)$, with $\dim{\cal E}_{p,q}=p>0$ and constant curvature
$q/p$. Geometrically, any such module is the space of sections
$\module_{p,q}=\Gamma(\torus^2,E_{p,q})$ of a complex vector bundle
$E_{p,q}\to\torus^2$ of rank $p$, Chern number $q$, and structure
group $U(p)$. The direct sum decompositions (\ref{directsumdecomp})
correspond to the usual Atiyah-Bott bundle splittings~\cite{AB}
\beq
E_{p,q}={\bigoplus}_k\,E_{p_k,q_k}
\label{bundlesplitting}\eeq
into sub-bundles $E_{p_k,q_k}\subset E_{p,q}$ about each Yang-Mills
critical point on $\torus^2$. The first two partition constraints in
(\ref{partitiondef}) for $\theta=0$ correspond to those on the rank of
(\ref{bundlesplitting}), $p=\sum_kp_k$ with $p_k>0$. This condition
coincides with (\ref{nuapartition}), where $\nu_a$ is the number of submodules
$\module_{p_k,q_k}$ of dimension $a$ (equivalently the number of
sub-bundles $E_{p_k,q_k}$ of rank $a$). The action (\ref{actionnuq})
is precisely of the form (\ref{NCYMpartition}) at $\theta=0$ and
without the background flux subtraction, while the exponential prefactors in
(\ref{Zptotalinst}) correspond to the fluctuation determinants
$W(\vp,\vq)$ in (\ref{Zpqlocclass}).

The third constraint in (\ref{partitiondef}) on the
magnetic charges $q_k$, which are dual to the lengths of the rows of
the Young tableaux of $U(p)$, restricts the gauge theory to a
particular isomorphism class of bundles over the torus. It is
straightforward to rewrite the partition function (\ref{Zptotalinst})
of {\it physical} Yang-Mills theory, defined as a weighted sum over
contributions from topologically distinct vector bundles over
$\torus^2$, in terms of that of Yang-Mills theory defined on a particular
isomorphism class $\module_{p,q}$ of projective modules over ${\cal
  A}_\theta$ up to irrelevant constants as
\beq
Z_p=\sum_{q=-\infty}^\infty(-1)^{(p-1)q}~Z_{p,q} \ ,
\label{Zpweighted}\eeq
where
\beq
Z_{p,q}=\sum_{\stackrel{\scriptstyle{\rm partitions}}
{\scriptstyle(\vp,\vq)}}(-1)^{|\vnu|}~\prod_{a=1}^p
\frac{\left(g^2a^3/2\pi^2\right)^{-\nu_a/2}}{\nu_a!}~
\e^{-S(\vp,\vq)} \ .
\label{Zpqcomm}\eeq
The partition sum here arises from the sum over cycle decompositions
that appears in the group theoretic setting above, with the number of
partition components $|\vnu|$ (or cycles) given by (\ref{vnudef}).

\subsection{The Exact Vacuum Amplitude}
\label{subsec:3.2}

{}From the commutative partition function (\ref{Zpqcomm}) we may now
extract the exact expression for the noncommutative field theory defined for
{\it any} $\theta$ in the following manner. We use the fact, reviewed
in section~5 of~\cite{ncproc1}, that gauge Morita equivalence provides
a one-to-one correspondence between projective modules over different
noncommutative tori (i.e. for different $\theta$'s) associated with
different topological numbers, augmented with transformations of
connections between the modules. It is an exact symmetry of the
noncommutative Yang-Mills action (\ref{YMactiondef}) which is firmly
believed to extend to the full quantum level. There are many good
pieces of evidence in support of this assumption~\cite{amns,SW,A-GB}.

In the present case, we will use the fact that Morita duality can be used
to map the quantum partition function of {\it ordinary} $\theta'=0$
Yang-Mills theory on $\torus^2$ onto noncommutative gauge theory with
deformation parameter $\theta=n/s$. The dimensionless coupling
constant and module dimensions in (\ref{Zpqcomm}) transform in this case as
$g^2=|s|^3\,g^{\prime\,2}$ and $\dim\module=\dim\module'/|s|$. The
equivalence provides a one-to-one correspondence between classical
solutions in the two field theories, i.e. their partitions. The
symmetry factors $\nu_a!$ in (\ref{Zpqcomm}) corresponding to
permutation of partition components of identical dimension are
preserved, as is the total number $|\vnu|$ of submodules in any given
partition $(\vp,\vq)$. From these facts it follows that the
fluctuation factors in (\ref{Zpqcomm}) are invariant under this Morita
duality {\it only} if the indices $a$ transform as
\beq
a=a'/|s| \ ,
\label{atransf}\eeq
which is equivalent to the expected requirement that the cycle lengths
$a$ be interpreted as the dimensions of submodules in the commutative
gauge theory.

With these identifications we can now straightforwardly map
(\ref{Zpqcomm}) onto the exact partition function of the $\theta=n/s$
Morita equivalent noncommutative gauge theory. The key point is that
the localization arguments which led to (\ref{Zpqlocclass}) do not
distinguish between the commutative, rational or irrational cases
there. All of the analysis and formulas of the previous section hold
universally for any value of $\theta$, and hence so should the exact
expression for the vacuum amplitude. Thus, given the generic
structure of partitions as outlined in section~\ref{subsec:1.3},
including the general definition of $\nu_a$, the final analytic
expression for the partition function of gauge theory on a fixed
projective module over the noncommutative torus, for any value of the
noncommutativity parameter $\theta$, is given by
\bea
Z_{p,q}&=&\sum_{\stackrel{\scriptstyle{\rm partitions}}
{\scriptstyle(\vp,\vq)}}~{\prod}_a\,\frac{(-1)^{\nu_a}}{\nu_a!}\,
\left(\frac{g^2}{2\pi^2}\,\big(p_a-q_a\theta\big)^3\right)^{-\nu_a/2}
\non&&\times\,\exp\left[-\frac{2\pi^2}{g^2}\,{\sum}_k\,
\left(p_k-q_k\theta\right)\left(\frac{q_k}{p_k-q_k\theta}-
\frac q{p-q\,\theta}\right)^2\,\right] \ .
\label{Zpqexactfinal}\eea
We have reinstated the constant curvature of $\module_{p,q}$, as it is
required to ensure that the Yang-Mills action transforms
homogeneously under Morita duality. This technique thereby explicitly
determines the fluctuation determinants $W(\vp,\vq)$ of the
semi-classical expansion (\ref{Zpqlocclass}).

We close this section with a brief description of how the expansion
(\ref{Zpqexactfinal}) elucidates the relations with and modifications
of ordinary Yang-Mills theory on the torus:
\begin{itemize}
\item It can be shown~\cite{inprep} that the partition function
  (\ref{Zpqexactfinal}) is a smooth function of $\theta$, even about
  $\theta=0$. At least at the level of two-dimensional noncommutative
  gauge theory, violations of $\theta$-smoothness in the quantum
  theory disappear at the non-perturbative level.
\item The Morita equivalence between rational noncommutative
  Yang-Mills theory on a projective module $\module_{p,q}$ with
  deformation parameter $\theta=n/s$, $n,s>0$ relatively prime, and
  ordinary non-abelian gauge theory is particularly transparent in
  this formalism. As mentioned above, for $\theta'=0$ the module
  dimensions transform as $\dim\module=\dim\module'/s$, and since in
  the commutative theory the bundle ranks are always positive
  integers, any module $\module$ in the rational theory has dimension
  bounded as $\dim\module\geq1/s$. Since $\dim\module_{p,q}=
  p-nq/s$, it follows that any partition $(\vp,\vq)$ of the rational
  theory consisting of submodules of dimension $\geq1/s$ has at most
  $\frac{p-nq/s}{1/s}=ps-qn$ components. Thus any gauge theory dual to
  this one admits partitions with $ps-qn$ components. In particular,
  as we have seen in the previous subsection, for $U(N)$ commutative
  Yang-Mills theory the maximum number of components is precisely the rank
  $N$, corresponding to the cycle decomposition with $\nu_1=N$ and
  $\nu_a=0~~\forall a>1$. Putting these facts together we arrive at
  the well-known result that noncommutative Yang-Mills theory with
  $\theta=n/s$ on a module $\module_{p,q}$ is Morita equivalent to
  $U(N)$ commutative gauge theory on $\torus^2$ with rank $N=ps-qn$.
\item The expansion (\ref{Zpqexactfinal}) clearly shows the
  differences between the commutative and noncommutative gauge
  theories. In the rational case $\theta=n/s$, all partitions contain
  at most $ps-qn$ submodules of $\module_{p,q}$ of dimension
  $\geq1/s$. But for $\theta$ irrational, there is no {\it a
  priori} bound on the number of submodules in a partition (although
  it is always finite) and submodules of arbitrarily small dimension
  can contribute to the partition function (\ref{Zpqexactfinal}). In
  particular, in this case we can approximate $\theta$ by a sequence
  of rational numbers, $\theta=\lim_mn_m/s_m$ with both $n_m,s_m\to\infty$ as
  $m\to\infty$. The rigorous way to take the limit of the
  noncommutative field theory is described in~\cite{lls1}. In the
  rational gauge theory with noncommutativity parameter
  $\theta_m=n_m/s_m$, the dimension of any submodule is bounded from
  below by $1/s_m$. It follows that any rational approximation to the
  vacuum amplitude $Z_{p,q}$ contains contributions from partitions of
  arbitrarily small dimension. Thus although formally similar, the
  exact expansion (\ref{Zpqexactfinal}) of the partition function has
  drastically different analytic properties in the commutative and
  noncommutative cases.
\end{itemize}

\section{Instanton Contributions}
\label{sec:4}

The fact that gauge theory on the noncommutative torus has an exact
semi-classical expansion in powers of $\e^{-1/g^2}$ suggests that it
should admit an interpretation in terms of non-perturbative
contributions from instantons of the
two-dimensional gauge theory. By an instanton we mean a finite action
solution of the Euclidean Yang-Mills equations (\ref{NCYMeom}) which
is not a gauge transformation of the trivial gauge field configuration
$A=0$. Interpreting (\ref{Zpqexactfinal}) in terms of such
configurations is not as straightforward as it may seem, because the
contributions to the sum as they stand are not arranged into gauge
equivalence classes. In this section we will briefly describe how to
rearrange the semi-classical expansion (\ref{Zpqexactfinal}) into a
sum over (unstable) instantons. This will entail a deep analysis of
the moduli spaces of the noncommutative gauge theory and will also
naturally motivate, via a comparison with corresponding structures on
the noncommutative plane, a matrix model analysis of the field theory
which will be carried out in the next section.

\subsection{Topological Yang-Mills Theory}
\label{subsec:4.1}

We will begin by studying the weak-coupling limit of the
noncommutative gauge theory as it is the simplest case to describe. In
the limit $g^2\to0$, the only non-vanishing contribution to
(\ref{Zpqexactfinal}) comes from those partitions for which the
Yang-Mills action attains its global minimum of~$0$. The only
partition for which this happens is the trivial one $(\vp,\vq)=(p,q)$
associated to the original Heisenberg module $\module_{p,q}$
itself. The corresponding moduli space of classical solutions
is the space of constant curvature connections on $\module_{p,q}$
modulo gauge transformations. Such classical configurations preserve
$\frac12$ of the supersymmetries in an appropriate supersymmetric
extension of the gauge theory~\cite{CDS,SchwarzMorita}. In this context, the
classical solutions live in a Higgs branch of the $\frac12$-BPS moduli space,
with the whole moduli space determined by a fibration over the Higgs
branch.

As described in detail in~\cite{ncproc1}, as a vector space the
Heisenberg module is given by
$\module_{p,q}=L^2(\real)\otimes\complex^q$, where $L^2(\real)$ is the
irreducible Schr\"odinger representation of the constant curvature
condition, and $\complex^q$ is the $q\times q$ representation of the
Weyl-'t~Hooft algebra in two dimensions. The latter algebra is known
to possess a unique irreducible unitary representation of dimension
$q/N$, $N={\rm gcd}(p,q)$, so that module decomposes into irreducible
components as
\beq
\module_{p,q}=L^2(\real)\otimes\left({\cal W}_{\zeta_1}\oplus\cdots
\oplus{\cal W}_{\zeta_N}\right) \ ,
\label{Epqdecomp}\eeq
where ${\cal W}_\zeta\subset\complex^q$ are the irreducible
representations of the Weyl-'t~Hooft algebra and
$\zeta\in\tilde\torus^2$ generate its center, with values in a dual
torus to the original one $\torus^2$. The only gauge transformations
which act trivially on (\ref{Epqdecomp}) are those which live in the
Weyl subgroup of $U(N)$, and dividing by this we find that the moduli
space of constant curvature connections on $\module_{p,q}$ is the
$N^{\rm th}$ symmetric product
\beq
{\cal M}_{p,q}={\rm Sym}^N\,\tilde\torus^2\equiv\left(\tilde
\torus^2\right)^N/\,S_N \ .
\label{Mpqconstcurv}\eeq
Remarkably, this space coincides with ${\rm
  Hom}(\pi_1(\torus^2),U(N))/U(N)$, the moduli space of {\it flat}
$U(N)$ bundles over the torus $\torus^2$ in commutative gauge theory~\cite{AB}.

Now let us examine more closely the partition function
(\ref{Zpqexactfinal}) in the limit $g^2\to0$. After using Morita duality to
remove the background flux contribution, the series receives
contributions only from partitions with vanishing magnetic charges
$q_k=0~~\forall k$, and we find
\beq
Z_{p,q}\big|_{g^2=0}=\sum_{\vnu\,:\,\sum_aa\nu_a=N}~\prod_{a=1}^N
\frac{(-1)^{\nu_a}}{\nu_a!}~\left(\frac{g^2a^3}{2\pi^2}
\right)^{-\nu_a/2}+O\left(\e^{-1/g^2}\right) \ .
\label{Zpqg0}\eeq
We thereby find that the weak coupling limit is independent of the
noncommutativity parameter $\theta$, and in particular it coincides
with the commutative version of the theory with structure group
$U(N)$. This is easiest to see from the form (\ref{auxphiint}), whose
$g^2=0$ limit gives explicitly
\beq
Z_{p,q}\big|_{g^2=0}=\int\dd\phi~\int\dd A~\dd\psi~\e^{-\ii\Tr(\psi
\wedge\psi-\phi\,F_A)} \ .
\label{ZpqTYM}\eeq
The integration over $\phi$, after reinstating the proper constant
curvature subtraction in (\ref{ZpqTYM}), localizes this functional
integral onto gauge field configurations of constant curvature, and the
partition function thereby computes the symplectic volume of
the moduli space (\ref{Mpqconstcurv}) with respect to the symplectic
structure on ${\cal M}_{p,q}$ inherited from the one (\ref{omegaab})
on ${\cal C}(\module_{p,q})$. It is formally the same as that of
topological Yang-Mills theory on $\torus^2$, except that now the
noncommutativity (through Morita equivalence) identifies
(\ref{Mpqconstcurv}) as the space of {\it all} constant curvature
connections, in contrast to the usual case where it only corresponds
to flat gauge connections.

In this case, the gauge theory is BRST equivalent (in the sense
described in section~\ref{subsec:2.3} for $t\to\infty$) to that with
gauge fixing functional
$V=\Tr\{\frac12\,(H-4F_A)+\nabla\lambda\cdot\psi\}$, where we have
introduced pairs $(\lambda,\eta)$ and $(\chi,H)$ of anti-ghost
multiplets of ghost numbers $(-2,-1)$ and $(-1,0)$, respectively, with
$\lambda,H$ bosonic and $\eta,\chi$ Grassmann-valued fields. Their
BRST transformation rules are
\bea
[{\sf Q}_\phi,\lambda]&=&\ii\eta \ , \non{}\{{\sf Q}_\phi,\eta\}&=&
[\phi,\lambda] \ , \non{}\{{\sf Q}_\phi,\chi\}&=&H \ , \non{}
[{\sf Q}_\phi,H]&=&\ii[\phi,\chi] \ ,
\label{XtraBRST}\eea
and the ${\sf Q}_\phi$-invariant action $S_{\rm
  D}\equiv-\ii\{{\sf Q}_\phi,V\}$ is given by
\bea
S_{\rm D}&=&\Tr\left\{\mbox{$\frac12$}\,
(H-F_A)^2-\mbox{$\frac12$}\,(F_A)^2-\ii\chi\,\nabla\wedge\psi+
\ii\nabla\eta\cdot\psi\right.\non&&+\left.\mbox{$\frac12$}\,
\chi\,[\chi,\phi]+\nabla\lambda\cdot\nabla\phi+\ii[\psi,\lambda]
\cdot\psi\right\} \ .
\label{SD2D}\eea
The functional $V$ conserves ghost number and the action (\ref{SD2D})
has non-degenerate kinetic energy, as in the case of the original
Yang-Mills system of section~\ref{subsec:2.3}. It gives the action of
two-dimensional Donaldson theory, and in this way the full
noncommutative gauge theory can be used to extract information about
the intersection pairings on the moduli space ${\cal
M}_{p,q}$~\cite{Wittensemi}.

Going back to the formula (\ref{Zpqg0}), we see that it involves a sum
over cycles $\vnu$ of terms which are singular at $g^2=0$. These terms
represent contributions to the symplectic volume from the conical
orbifold singularities of the moduli space (\ref{Mpqconstcurv}), which arise
due to the existence of reducible connections. For
this, we note that the fixed point locus of a conjugacy class element
$\sigma\in[1^{\nu_1}\,2^{\nu_2}\cdots p^{\nu_p}]$ acting on
$(\zeta_1,\dots,\zeta_N)\in(\tilde\torus^2)^N$ is
$\prod_a(\tilde\torus^2)^{\nu_a}$. The action of the corresponding
stabilizer subgroup of $S_N$ is
$\prod_aS_{\nu_a}\ltimes(\zed_a)^{\nu_a}$, where the symmetric group
$S_{\nu_a}$ permutes coordinates in the factor
$(\tilde\torus^2)^{\nu_a}$ while the cyclic group $\zed_a$ acts in
each cycle of length $a$. Only the $S_{\nu_a}$ factors act
non-trivially, and so the singular locus of ${\cal M}_{p,q}$ is a
disjoint union over the conjugacy classes $[1^{\nu_1}\,2^{\nu_2}\cdots
p^{\nu_p}]\subset S_N$ of the strata $\prod_a{\rm
  Sym}^{\nu_a}\,\tilde\torus^2$, as reflected by the expansion
(\ref{Zpqg0}).

\subsection{Instanton Partitions}
\label{subsec:4.2}

Let us now consider the general case. The basic problem is that there
is an isomorphism $\module_{mp,mq}\cong\oplus^m\,\module_{p,q}$ of
Heisenberg modules, owing to the reducibility of the Weyl-'t~Hooft
algebra, with $\module_{mp,mq}$ and $\module_{p,q}$ both possessing
the {\it same} constant curvature. We circumvent this problem by
writing each component of a given partition as
$(p_k,q_k)=N_k(p_k',q_k')$, with $N_k={\rm gcd}(p_k,q_k)$ and $p_k',q_k'$
relatively prime, and restrict the sum over partitions $(\vp,\vq)$
to those with {\it distinct} K-theory charges $(p_k',q_k')$. We call
such partitions ``instanton partitions''~\cite{inprep}, as they each represent
distinct, gauge equivalence classes of classical solutions to the
noncommutative Yang-Mills equations. Then the direct sum decomposition
(\ref{directsumdecomp}) is modified to
\beq
\module_{p,q}={\bigoplus}_a\,\module_{N_ap_a',N_aq_a'} \ ,
\label{decompmod}\eeq
and the corresponding moduli space of classical solutions is~\cite{inprep}
\beq
{\cal M}_{p,q}'={\prod}_a\,{\cal M}_{N_ap_a',N_aq_a'}=
{\prod}_a\,{\rm Sym}^{N_a}\,\tilde\torus^2 \ .
\label{modspinst}\eeq
The orbifold singularities present in (\ref{modspinst}) can now be
used to systematically construct the gauge inequivalent contributions
to noncommutative Yang-Mills theory. In this way one may rewrite the
expansion (\ref{Zpqexactfinal}) as a sum over instantons along with a
finite number of quantum fluctuations about each instanton,
representing a finite, non-trivial perturbative expansion in
$1/g$. For more details, see~\cite{inprep}.

\subsection{Fluxon Contributions}
\label{subsec:4.3}

The instanton solutions that we have found for gauge theory on the
noncommutative torus bear a surprising relationship to
soliton solutions of gauge theory on the noncommutative {\it
  plane}~\cite{poly}--\cite{gn}. The classical solutions of the noncommutative
Yang-Mills equations in this latter case are labelled by two integers, the rank
of the gauge group and the magnetic charge, similarly to the case of
the torus. These noncommutative solitons are termed ``fluxons'' and
they are finite energy instanton solutions which carry quantized
magnetic flux. The classical action evaluated on a fluxon of charge
$q$ is given by~\cite{gn}
\beq
S(q)=\frac{2\pi^2q}{g^2\theta} \ .
\label{fluxonaction}\eeq
This action is very similar to
(\ref{NCYMpartition}) in the limit $g^2\theta\to\infty$, and
in~\cite{gsv} it was described how to map the instanton expansion on
the noncommutative torus to one on the noncommutative plane by using Morita
equivalence and taking a suitable large area limit. In terms of the partition
sum
(\ref{Zpqexactfinal}), a fluxon of charge $q$ is composed of $\nu_a$
elementary vortices of charges $a=1,2,\dots$. The symmetry factors
$\nu_a!$ appear in (\ref{Zpqexactfinal}) to account for the fact that
vortices of equal charge inside the fluxon are identical, while the
moduli dependence (through the vortex positions) is accompanied by the
anticipated exponent $|\vnu|$, the total number of elementary vortex
constituents of the fluxon. The remaining terms correspond to quantum
fluctuations about each fluxon in the following manner.

The basic fluxon solution corresponds to the elementary vortex
configuration $\nu_1=q$, $\nu_a=0~~\forall a>1$. In the large area
limit, the semi-classical expansion (\ref{Zpqexactfinal}) can be
interpreted in terms of the contributions from basic fluxons of charge
$q$ and classical action (\ref{fluxonaction}), along with fluctuations
around the soliton solution, leading to the partition function~\cite{gsv}
\beq
{\cal Z}_q=\frac{\e^{-2\pi^2q/g^2\theta}}{N\,\sqrt{g^2\theta}}~
\sum_{\vnu\,:\,\sum_aa\nu_a=q}~\prod_{a=1}^q\left(-\frac1{\nu_a!}~
\sqrt{\frac{2\pi^2}{a^3g^2\theta^3}}~\right)^{\nu_a} \ .
\label{calZq}\eeq
The (unweighted) sum over topological charges can be performed
exactly and the result is
\beq
{\cal Z}\equiv\sum_{q=0}^\infty{\cal Z}_q=\exp\left[-\frac{
2\pi~\e^{-2\pi^2/g^2\theta}}{\sqrt{g^2\theta^3}}~\Phi\left(
\e^{-2\pi^2/g^2\theta}\,;\,\mbox{$\frac32$}\,;\,1\right)\right] \ ,
\label{calZqsum}\eeq
where the function
\beq
\Phi(z;s;\mu)=\sum_{k=0}^\infty\frac{z^k}{(k+\mu)^s}
\label{Phifndef}\eeq
is analytic in $z\in\complex$ with a branch cut from $z=1$ to
$z=\infty$. The instanton series has been resummed in (\ref{calZqsum})
into the non-perturbative exponential, which is typical of a dilute
instanton gas. This is not surprising, given that fluxons are
non-interacting objects and thereby lead to an extensive partition
function. It would be interesting to examine the dynamics of all the
instantons described in this picture on the moduli spaces
(\ref{Mpqconstcurv}) and (\ref{modspinst}), using the K\"ahler
structure inherited from the symplectic structure (\ref{omegaab}) and
metric $\Tr\alpha\cdot\beta$ on the space $\cal C$ of compatible gauge
connections.

The non-trivial results obtained for the noncommutative plane suggest
another way of tackling two-dimensional noncommutative gauge theory in
general~\cite{gn}. Since the planar algebra of functions is generated by the
coordinate operators $x^1,x^2$ obeying the Heisenberg algebra
$[x^1,x^2]=\ii\theta$, gauge connections act by inner automorphisms
and may be written as
\beq
D_i=\mbox{$\frac\ii\theta$}\,\epsilon_{ij}\,x^j+A_i
\label{DiNCplane}\eeq
for $i=1,2$. The curvature is given by
\beq
F_A=[D_1,D_2]-\mbox{$\frac\ii\theta$} \ ,
\label{FANCplane}\eeq
and after a rescaling of fields the partition function is defined by the
infinite dimensional matrix model
\beq
{\cal Z}=\lim_{\varepsilon\to0^+}\,\int\dd D_1~\dd D_2~
\exp\left[-\frac{\pi\,\theta}{2g^2}\,\Tr\big([D_1,D_2]-1\big)^2-
\varepsilon\,\Tr D\cdot D\right] \ .
\label{calZinfinite}\eeq
The second term in the action of (\ref{calZinfinite})
regulates the partition function and is a gauge-invariant analog of
the infrared regularization provided by the area of the torus. It is
required to ensure that the semi-classical approximation to the
functional integral exists. The classical fluxon solutions are
unstable critical points whose moduli are the positions of the
vortices~\cite{gn}. The Yang-Mills energy density of the vortices is
independent of these positions and integrating along these moduli would lead to
a
divergent path integral in (\ref{calZinfinite}). While this may seem
like a fruitful line of attack, it presents many
difficulties. Foremost among these is the fact that finite action
configurations would require the field strength $F_A$ to be a compact
operator. Since there are no bounded operators $D_i$ for which
(\ref{FANCplane}) is compact, the effective gauge configuration space
consists only of {\it unbounded} operators and the partition function
(\ref{calZinfinite}) is not naturally realized as the large $N$ limit
of a finite dimensional matrix model. This makes an exact solution
intractable. In the next section we shall present a matrix model
formulation of noncommutative gauge theory in two dimensions which
circumvents these difficulties, and enables an explicit and exact
analysis of the configurations described here.

\section{Combinatorial Quantization}
\label{sec:5}

In this final section we will show how a combinatorial approach can be
used to explicitly compute the partition function of noncommutative
gauge theory in two dimensions. Part of the motivation for doing this
was explained at the end of the last section. Another reason is to
make sense of the Feynman path integral over the space $\cal C$ of
compatible connections. We will approximate $\cal C$ by a
finite-dimensional $N\times N$ matrix group and then analyse the
partition function in the limit $N\to\infty$. The hope is then that
this procedure yields a concrete, non-perturbative definition of the
noncommutative field theory. This matrix model is intimately connected
with a lattice regularization of the noncommutative gauge theory
obtained by triangulating $\torus^2$, and restricting to modules over
the finite-dimensional matrix algebras. In this setting the
non-trivial K-theory of the torus algebra ${\cal A}_\theta$ is lost,
and as in section~\ref{subsec:3.1} the computation will give the
Yang-Mills partition function summed over all topological types of
projective modules over ${\cal A}_\theta$. We will begin by recalling
some salient features of commutative lattice gauge theory, and
contrast it with what happens in the noncommutative setting. Then we
will proceed to define and completely solve the discrete version of
noncommutative gauge theory in two dimensions, and describe how it can
be used to extract information about the continuum field theory. The
material contained in this section is new and presents a novel
explicit solution of noncommutative Yang-Mills theory.

\subsection{The Local Lattice Regularization}
\label{subsec:5.1}

\begin{figure}
\centering
\includegraphics[height=3cm]{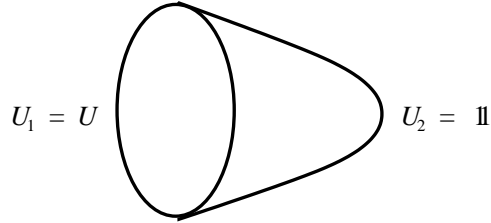}
\caption{The disk amplitude.}
\label{disk}\end{figure}

\begin{figure}
\centering
\includegraphics[height=3cm]{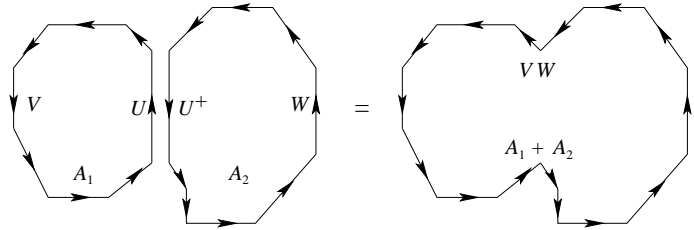}
\caption{The partition function of two-dimensional Yang-Mills theory
  is invariant under subdivision of the plaquettes of the lattice.}
\label{plaquette}\end{figure}

In ordinary two-dimensional Yang-Mills theory, the lattice form of the
quantum field theory~\cite{Wilson} possesses some very special properties and
provides an indispensible tool for obtaining its complete analytic
solution~\cite{Migdal,Rusakov}. Let us consider the partition function on a
disk of area $A$~(fig.~\ref{disk}). It can be obtained from the cylinder
amplitude
(\ref{cylampl}) by pinching the right boundary of the cylinder to a
point, so that $U_2$ becomes the holonomy surrounding a disk of
vanishing area. The corresponding physical state wavefunction is the
delta-function supported at the identity element $U_2=\id$ of the
unitary group with respect to its Haar measure,
$\Psi[U_2]=\delta(U_2,\id)$. Then from (\ref{cylampl}) with $U_1=U$
and $U_2=\id$ we obtain the disk amplitude
\beq
Z(A,U)={\sum}_R\,\dim R~\chi^{~}_R(U)~\e^{-\frac{g^2A}2\,C_2(R)} \ .
\label{diskampl}\eeq
By using the area-preserving diffeomorphism invariance of the theory,
we may interpret (\ref{diskampl}) as an amplitude for a {\it
  plaquette}, i.e. the interior of a simplex in a local triangulation
of the spacetime~(fig.~\ref{plaquette}).

One of the main advantages of the discrete formalism is its
self-similarity property~\cite{Migdal}. Consider the gluing together of two
disk amplitudes along a plaquette link as depicted in
fig.~\ref{plaquette}. The gluing property follows from
(\ref{diskampl}) and the fusion rule (\ref{fusionrule}) for the
characters, which together imply
\beq
\int[\dd U]~Z(A_1,VU)\,Z(A_2,U^\dag W)=Z(A_1+A_2,VW) \ .
\label{gluingrule}\eeq
This result expresses the renormalization group invariance of the
basic plaquette Boltzmann weight. Subdivision of the lattice into a
very fine lattice yields a result which converges to that in the
continuum theory. But (\ref{gluingrule}) implies that the computation
may be carried out on an arbitrarily coarse lattice. Hence the
lattice field theory produces the exact answer and the continuum limit
is trivial. From this treatment it is in fact possible to directly
obtain the torus amplitude (\ref{torusamplgroup}).

As we will soon see, this self-similarity property under gluing of
plaquettes is not shared by the noncommutative version of the lattice
gauge theory, reflecting its inherent non-locality. Noncommutativity
introduces long-ranged interactions between plaquettes of the
lattice. A clear way to understand this breakdown is to recall the
Gross-Witten reduction of $U(N)$ Yang-Mills theory on
$\real^2$~\cite{GrossWitten}. The calculation of the lattice partition function
in this case can be
easily reduced to a single unitary matrix integration by exploiting
the gauge invariance of the theory. This is possible to do by fixing
an axial gauge and an appropriate change of variables. If $U_i(x)$ denotes the
operator of parallel transport from a lattice site $x$ to its
neighbouring point along a link in direction $\hat\imath$, $i=1,2$,
then one may fix the gauge $U_1(x)=\id~~\forall x$. This renders the
theory trivial in the $\hat1$ direction. There is a residual gauge
symmetry which may be used to define $U_2(x+\hat1)=W(x)\,U_2(x)$, and
the partition function thereby factorizes into a product of decoupled
integrals over the unitary matrices $W(x)$~\cite{GrossWitten}. This is {\it
not} possible to do in the noncommutative gauge theory, because in its lattice
incarnation it is required to be formulated on a periodic lattice as a
result of UV/IR mixing~\cite{amns}, and large gauge transformations thereby
forbid axial gauge choices. As expected, UV/IR mixing drastically alters the
Wilsonian renormalization features of the noncommutative field theory,
and it admits non-trivial scaling limits. Later on we will see how
noncommutativity explicitly modifies the Gross-Witten result.

\subsection{Noncommutative Lattice Gauge Theory}
\label{subsec:5.2}

We will now proceed to formulate and explicitly solve the
noncommutative version of lattice gauge theory, which gives yet
another proof of the exact solvability of the continuum theory. We
discretize the torus of the previous sections as an $L\times L$ periodic
square lattice. For convenience, we assume that $L$ is an odd integer. Let
$\varepsilon$ be the dimensionful lattice spacing, so that the area of the
discrete torus is
\beq
A=\varepsilon^2L^2 \ .
\label{extent}\eeq
Any function $f(x)$ on the periodic lattice admits a Fourier series
expansion over a Brillouin zone $\zed_L\times\zed_L$,
\beq
f(x)=\frac1{L^2}\,\sum_{\vm\in(\zed_L)^2}f^{~}_\vm
{}~\e^{2\pi\ii m_i\,x^i/\varepsilon L} \ .
\label{Fourierlat}\eeq
A natural lattice star-product may be defined as the proper discretized version
of the integral kernel representation of the continuum star product
as
\beq
\left(f\star g\right)(x)=\frac1{L^2}\,{\sum}_{y,z}\,f(x+y)\,g(x+z)~
\e^{2\pi\ii y\wedge z/\varepsilon^2L} \ ,
\label{latticestardef}\eeq
where the sums run over lattice points. This identifies $\theta=2/L$
and hence the dimensionful noncommutativity parameter of the commutant
algebra as
\beq
\Theta=\frac{\theta A}{2\pi}=\frac{\varepsilon^2L}\pi \ .
\label{thetalat}\eeq
As mentioned in the previous subsection, because of a kinematical
version of UV/IR mixing, the lattice regularization of noncommutative
field theory {\it requires} the space to be a torus~\cite{amns}.

We will now write down, in analogy with the commutative case, the natural,
nonperturbative lattice regularization of the continuum noncommutative
gauge theory. This is provided by the noncommutative
version of the standard Wilson plaquette model~\cite{Wilson}. The partition
function is~\cite{amns}
\beq
{\cal Z}_r=\int{\prod}_x\,\big[\dd U_1(x)\big]~\big[
\dd U_2(x)\big]~\exp\left[\frac1{4\lambda^2L}\,
{\sum}_\Box\,\tr^{~}_N\left(U^{~}_\Box+U_\Box^\dag\right)\right] \ ,
\label{calZdef}\eeq
where
\beq
\lambda=\sqrt{g^2\varepsilon^2L}
\label{tHooftcoupling}\eeq
is the 't~Hooft coupling constant. Here
$\prod_x[\dd U_i(x)]$ is the normalized, invariant Haar
measure on the ordinary $r\times r$ unitary group $U(r)$ with
\beq
r=L\cdot N \ ,
\label{rLN}\eeq
and $N$ is the rank of the given module. The fields $U_i(x)$ are $U(N)$ gauge
fields which live at the links $(x,i)$ of the lattice and which are
``star-unitary'',
\beq
(U_i\star U_i^\dag)(x)=(U_i^\dag\star U_i)(x)=\id_r \ .
\label{starunitary}\eeq
In the continuum limit $\varepsilon\to0$, they are identified with the gauge
fields of the previous sections through $U_i=\e^{\star\,\varepsilon A_i}$. The
sum in (\ref{calZdef}) runs through the plaquettes $\Box$ of the lattice with
$U^{~}_\Box$ the ordered star-product of gauge fields around the plaquette,
\beq
U^{~}_\Box=U_1(x)\star U_2(x+\varepsilon
\,\hat1)\star U_1(x+\varepsilon\,\hat2)^\dag\star U_2(x)^\dag \ ,
\label{UBox}\eeq
where $x$ is the basepoint of the plaquette and $\hat\imath$ denotes the unit
vector along the $i^{\,\rm th}$ direction of the lattice. The lattice
gauge theory (\ref{calZdef}) is invariant under the gauge transformation
\beq
U_i(x)~\longmapsto~g(x)\star U_i(x)\star g(x+\varepsilon\,\hat\imath)^\dag \ ,
\label{gaugetransflat}\eeq
where the gauge function $g(x)$ is star-unitary,
\beq
(g\star g^\dag)(x)=(g^\dag\star g)(x)=\id_r \ .
\label{gaugestarunitary}\eeq

\subsection{Gauge Theory on the Fuzzy Torus}
\label{subsec:5.3}

The feature which makes the noncommutative lattice gauge theory (\ref{calZdef})
exactly solvable is that the entire lattice formalism presented above can be
cast into a finite dimensional version of the abstract algebraic description of
gauge theory on a projective module over the noncommutative
torus~\cite{AMNSFinite}. For this, we note that since the noncommutativity
parameter of the
commutant algebra is the rational number $\theta=2/L$, the generators
$Z_i$ of $\alg_{\theta}$ obey the commutation relations
\beq
Z_1\,Z_2=\e^{4\pi\ii/L}~Z_2\,Z_1 \ .
\label{Zprimealg}\eeq
This algebra admits a finite dimensional representation which gives the
noncommutative space the geometry of a {\it fuzzy} torus. Namely,
$\alg_{\theta}$ can be represented on the finite dimensional Hilbert module
$\complex^L$, regarded as the space of functions on the finite cyclic group
$\zed_L$, as
\beq
Z_1=V_L \ , ~~ Z_2=\left(W_L\right)^2 \ ,
\label{Zprimerep}\eeq
where $V_L$ and $W_L$ are the $SU(L)$ shift and clock matrices which
obey $V_L\,W_L=\e^{2\pi\ii/L}~W_L\,V_L$.

Since $(Z_i)^L=\id_L$, the matrices (\ref{Zprimerep}) generate
the finite-dimensional algebra $\mat_L$
of $L\times L$ complex matrices. In fact, they provide a one-to-one
correspondence between lattice fields (\ref{Fourierlat}) with the star-product
(\ref{latticestardef}) and $L\times L$ matrices through
\beq
\hat f=\frac1{L^2}\,\sum_{\vm\in(\zeds_L)^2}f^{~}_\vm
{}~\e^{-2\pi\ii m_1m_2/L}~Z_1^{m_1}\,Z_2^{m_2} \ .
\label{hatffinite}\eeq
It is easy to check that this correspondence possesses the same formal
properties as in the continuum, namely
\bea
\tr^{~}_L~\hat f&=&f^{~}_{\mbf 0}=\frac1{L^2}\,{\sum}_x\,f(x) \ ,
\label{matrixfieldprop1}\\\hat f\,\hat g&=&\widehat{f\star g} \ .
\label{matrixfieldprop2}\eea
In particular, the star-unitarity condition (\ref{starunitary}) translates into
the requirement
\beq
\hat U^{~}_i\,\hat U_i^\dag=\hat U_i^\dag\,\hat U^{~}_i=\id_r \ .
\label{hatUiunitary}\eeq
Therefore, there is a one-to-one correspondence between $N\times N$
star-unitary matrix fields $U_i(x)$ and $r\times r$ unitary matrices $\hat
U_i$. In the parlance of the geometry of the noncommutative torus, we have
$\alg_{\theta}\cong\mat_L$ and the endomorphism algebra is ${\rm
End}({\cal E})\cong\alg_{\theta}\otimes\mat_N\cong\mat_r$. The
gauge fields in the present setting live in the unitary group of this
algebra, which is just $U(r)$ as above.

To cast the gauge theory (\ref{calZdef}) into a form which is the natural
nonperturbative version of (\ref{NCYMpartitiondef})~\cite{AMNSFinite}, we
introduce connections $V_i=\e^{\varepsilon\,\nabla_i}$ on this discrete
geometry which are $r\times r$ unitary matrices that may be decomposed in terms
of the gauge fields $\hat U_i$ as
\beq
V_i=\hat U_i\,\Gamma_i \ ,
\label{hatVUGamma}\eeq
where the matrices $\Gamma_i=\e^{\varepsilon\,\partial_i}$
correspond to lattice shift operators. They thereby satisfy the
commutation relations
\bea
\Gamma_1\,\Gamma_2&=&\zeta~\Gamma_2\,\Gamma_1 \ , \label{HWalg}\\
\Gamma_i\,Z_j\,\Gamma_i^\dag&=&\e^{2\pi\ii\delta_{ij}/L}~Z_j \ ,
\label{GammaZprime}\eea
where
\beq
\zeta=\e^{2\pi\ii q/L}
\label{zetadef}\eeq
is a $\zed_L$-valued phase factor whose continuum limit gives the
background flux in (\ref{YMactiondef}). The integer $q$ is chosen,
along with some other integer $c$, to satisfy the Diophantine equation
\beq
cL-2q=1
\label{Dioeq}\eeq
for the relatively prime pair of integers $(L,2)$. The equations (\ref{HWalg})
and (\ref{GammaZprime}) can then be solved by
\beq
\Gamma_1=\left(W_L^\dag\right)^{2q} \ , ~~
\Gamma_2=\Bigl(V_L\Bigr)^q \ .
\label{Gammasol}\eeq
Note that while the Heisenberg commutation relations for constant
curvature connections admit no
finite dimensional representations, the Weyl-'t~Hooft commutation relation
(\ref{HWalg}), which is its exponentiated version, does. In other words, the
matrices (\ref{Gammasol}) generate the irreducible action of the
Heisenberg-Weyl group on the finite-dimensional algebra
$\alg_{\theta}\cong\mat_L$. This construction can be generalized to
provide discrete versions of the standard Heisenberg modules over the
noncommutative torus~\cite{AMNSFinite}.

We now substitute the matrix-field correspondences (\ref{Fourierlat}) and
(\ref{hatffinite})--(\ref{matrixfieldprop2}) for the gauge fields into the
partition function (\ref{calZdef}), use the fact that $\Gamma_i$
generates a lattice shift along direction $\hat\imath$, and use the
decomposition (\ref{hatVUGamma}) to rewrite the action in terms of the
finite dimensional connections $V_i$. By using in addition the
invariance of the Haar measure, the Weyl-'t~Hooft algebra
(\ref{HWalg}), and the representation of the trace
$\tr^{~}_r=\tr^{~}_L\otimes\tr^{~}_N$ on ${\rm
  End}({\cal E})\cong\mat_r$, after some algebra we find
that the partition function (\ref{calZdef}) can be written finally as
the unitary two-matrix model~\cite{AMNSFinite}
\beq
{\cal Z}_r=\int[\dd V_1]
{}~[\dd V_2]~\e^{\frac1{2\lambda^2}~{\rm Re}\,\tr^{~}_r\,
\zeta~V_1\,V_2\,V_1^\dag\,V_2^\dag} \ .
\label{calZTEK}\eeq
This is the partition function of the twisted Eguchi-Kawai model in
two dimensions~\cite{EK,EKtwist}, with twist given by the $\zed_L$ phase factor
(\ref{zetadef}), and it coincides with the dimensional reduction of
ordinary Wilson lattice gauge theory to a single plaquette~\cite{amns}. The
star-gauge invariance (\ref{gaugetransflat}) of the plaquette model
(\ref{calZdef}) corresponds to the $U(r)$ invariance
\beq
V_i~\longmapsto~\hat g\,V_i\,\hat g^\dag \ , ~~ \hat g\in U(r)
\label{TEKinv}\eeq
of the twisted Eguchi-Kawai model (\ref{calZTEK}). Note that the $U(r)$ gauge
symmetry of the matrix model (\ref{calZTEK}) is a mixture of the original
$L\times L$ spacetime degrees of freedom of the noncommutative lattice
gauge theory (\ref{calZdef}) and its $U(N)$ colour symmetry. The partition
function (\ref{calZTEK}) is a well-defined, finite-dimensional operator version
of the noncommutative Wilson lattice gauge theory in two-dimensions, which we
will now proceed to compute explicitly.

\subsection{Exact Solution}
\label{subsec:5.4}

To evaluate the unitary group integrals (\ref{calZTEK}), we insert an extra
integration involving the gauge invariant delta-function acting on class
functions on $U(r)$ to get
\beq
{\cal Z}_r=\int[\dd V_1]~[\dd V_2]~
\int[\dd W]~\delta\left(W\,,\,V_1\,V_2\,
V_1^\dag\,V_2^\dag\right)~\e^{\frac1{4\lambda^2}\,
\tr^{~}_r(\zeta\,W+\overline{\zeta}\,W^\dag)} \ .
\label{ZTEKdelta}\eeq
The delta-function in the Haar measure may be expanded in terms of the
orthornormal $U(r)$ characters as
\beq
\delta(W,U)={\sum}_R\,\chi^{~}_R(W)\,\chi^{~}_R(U^\dag) \ .
\label{deltachi}\eeq
As in section~\ref{subsec:3.1}, the unitary irreducible
representations $R$ of the Lie group $U(r)$ may be parametrized by
partitions $\vn=(n_1,\dots,n_r)$ into $r$ parts of decreasing integers
as in (\ref{ndecrease}). The character of the unitary matrix $W$
in this representation can then be written explicitly by means of the
Weyl formula
\beq
\chi^{~}_R(W)=\chi^{~}_\vn(W)=\frac{
\det_{a,b}\left[\e^{\ii(n_a-b+r)\phi_b}\right]}
{\det_{a,b}\left[\e^{\ii(a-1)\phi_b}\right]} \ ,
\label{Weylformula}\eeq
where $\e^{\ii\phi_1},\dots,\e^{\ii\phi_r}$ are the eigenvalues of $W$.

On substituting (\ref{deltachi}) into (\ref{ZTEKdelta}), the integration over
$V_1$ and $V_2$ can be carried out explicitly by using the fusion rule
(\ref{fusionrule}) for the $U(r)$ characters along with the fission relation
\beq
\int[\dd U]~\chi^{~}_\vn\left(U\,V\,
U^\dag\,W\right)=\frac{\chi^{~}_\vn(V)\,\chi^{~}_\vn(W)}{d_\vn} \ ,
\label{fission}\eeq
where
\beq
d_\vn=\chi^{~}_\vn(\id_r)=
\prod_{a<b}\left(1+\frac{n_a-n_b}{b-a}\right)
\label{dimirrep}\eeq
is the dimension $\dim R$ of the representation $R$ with highest weight
vector $\vn=(n_1,\dots,n_r)$. In this way the
partition function takes the form
\beq
{\cal Z}_r=\sum_{n_1>\dots>n_r}\,\frac1{d_\vn}~
\int[\dd W]~\chi^{~}_\vn(W)
{}~\e^{\frac1{4\lambda^2}\,\tr^{~}_r(\zeta\,W+
\overline{\zeta}\,W^\dag)} \ .
\label{calZsumint}\eeq
The twist factors (\ref{zetadef}) can be decoupled from the integration in
(\ref{calZsumint}) by the rescaling $W\to\overline{\zeta}\,W$ and by
using $U(r)$ invariance of the Haar measure along with the character identity
\beq
\chi^{~}_\vn\left(\,\overline{\zeta}\,W\right)=
\e^{-2\pi\ii q\,C_1(\vn)/L}~\chi^{~}_\vn(W) \ ,
\label{chiscale}\eeq
where
\beq
C_1(R)=C_1(\vn)=\sum_{a=1}^rn_a
\label{C1}\eeq
is the linear Casimir invariant of the representation $R$ which
counts the total number of boxes in the corresponding $U(r)$ Young tableau.

We now expand the invariant function in (\ref{calZsumint}) which
after rescaling is the Boltzmann factor for the one-plaquette $U(r)$ Wilson
action. Its character expansion can be given explicitly in terms of
modified Bessel functions $I_n(z)$ of the first kind of integer
order~$n$ which are defined by their generating function as
\beq
\exp\left[\frac z2\,\left(t+\frac1t\right)\right]=\sum_{n=-\infty}^\infty
I_n(z)~t^n \ .
\label{Ingenfn}\eeq
By using (\ref{Weylformula}) one finds~\cite{ID}
\beq
\e^{\beta\,\tr^{~}_r(W+W^\dag
)}=\sum_{n_1>\dots>n_r}\,\det_{a,b}\,\big[I_{n_a-a+b}
(2\beta)\big]~\chi^{~}_\vn\left(W^\dag\right) \ ,
\label{Wilsonchi}\eeq
and, by using the fusion rule (\ref{fusionrule}), substitution of
(\ref{Wilsonchi}) into (\ref{calZsumint}) gives a representation of
the lattice partition function as a sum over a single set of
partitions alone,
\beq
{\cal Z}_r=\sum_{n_1>\dots>n_r}\frac{\e^{-2\pi\ii q\,C_1(\vn)/L}}
{d_\vn}~\det_{a,b}\,\big[I_{n_a-a+b}(1/2\lambda^2)\big] \ .
\label{calZpartitions}\eeq

To express (\ref{calZpartitions}) as a perturbation series in the effective
coupling constant $1/\lambda^2$, we substitute into this
expression the power series expansion of the modified Bessel functions,
\beq
I_\nu(z)=\sum_{m=0}^\infty\,\frac1{m!~\Gamma(\nu+m+1)}\,\left(\frac z2
\right)^{\nu+2m} \ ,
\label{Besselseries}\eeq
where $\Gamma(z)$ is the Euler function. The infinite sum may then be extracted
out line by line from the determinant in (\ref{calZpartitions}) by using the
multilinearity of the determinant as a function of its $r$ rows, and we find
\bea
{\cal Z}_r&=&\sum_{n_1>\dots>n_r}\frac{\e^{-2\pi\ii q\,C_1(\vn)/L}}
{d_\vn}\,\sum_{m_1=0}^\infty\dots\sum_{m_r=0}^\infty~
\prod_{s=1}^r\frac{(1/2\lambda)^{2m_s}}{m_s!}
\nn\\&&\times\,\det_{a,b}\,\left[\frac{(1/2\lambda)^{2(m_a+n_a-a+b)}}
{\Gamma(m_a+n_a-a+b+1)}\right] \ .
\label{calZBesselseries}\eea
Note that the total contribution to (\ref{calZBesselseries}) vanishes from any
set of integers for which $m_a+n_a<a-r$ for any {\it single} index
$a=1,\dots,r$.

The determinant in (\ref{calZBesselseries}) can be evaluated explicitly as
follows. For any sequence of integers $s_1,\dots,s_r$, we have
\beq
\det_{a,b}\,\left[\frac{z^{s_a-a+b}}{\Gamma(s_a-a+b+1)}\right]
=z^{s_1+\dots+s_r}~\left|\begin{matrix}\frac1{\Gamma(s_1+1)}&
\frac1{\Gamma(s_2)}&\cdots&\frac1{\Gamma(s_r-r+2)}\\
\frac1{\Gamma(s_1+2)}&\frac1{\Gamma(s_2+1)}&\cdots&\frac1{\Gamma(s_r-r+3)}
\\\vdots&\vdots&\ddots&\vdots\\\frac1{\Gamma(s_1+r)}&
\frac1{\Gamma(s_2+r-1)}&\cdots&\frac1{\Gamma(s_r+1)}\end{matrix}\right| \ .
\label{detzGamma}\eeq
Factorizing $1/\Gamma(s_b-b+r+1)$ from each column $b$ of the remaining
determinant in (\ref{detzGamma}) yields
\bea
&&\det_{a,b}\,\left[\frac{z^{s_a-a+b}}{\Gamma(s_a-a+b+1)}\right]=
z^{s_1+\dots+s_r}~\prod_{b'=1}^r\frac1{\Gamma(s_{b'}-b'+r+1)}\nn\\&&
{}~~~~~~~~~~\times\,
\det_{a,b}\,\big[(s_b-b+a+1)(s_b-b+a+2)\cdots(s_b-b+r)\big] \ .
\label{detinter}\eea
The argument of the determinant in the right-hand side of
(\ref{detinter}) is a monic polynomial in
the variable $\alpha_b=s_b-b$ with highest degree term
$\alpha_b^{r-a}$. By using multilinearity of the determinant, it becomes
$\det_{a,b}\,[\alpha_a^{r-b}]=\prod_{a<b}(\alpha_a-\alpha_b)$, and we
arrive finally at
\bea
\det_{a,b}\,\left[\frac{z^{s_a-a+b}}{\Gamma(s_a-a+b+1)}\right]&=&
z^{s_1+\dots+s_r}~\prod_{b=1}^r\frac{(r-b)!}{\Gamma(s_b-b+r+1)}\nn\\&&
\times\,\prod_{a<b}\left(1+\frac{s_a-s_b}{b-a}\right) \ .
\label{calZdetgen}\eea
Note that if ${\mbf s}=(s_1,\dots,s_r)$ is a partition, then the last
product in (\ref{calZdetgen}) is just the dimension $d_{\mbf s}$ of the
corresponding $U(r)$ representation.

The partition function (\ref{calZBesselseries}) is thereby given as
\bea
{\cal Z}_r&=&\sum_{n_1>\dots>n_r}\,\frac{\e^{-2\pi\ii q\,
C_1(\vn)/L}}{d_\vn\,(2\lambda)^{2C_1(\vn)}}\nn\\&&\times\,
\sum_{m_1=0}^\infty\cdots
\sum_{m_r=0}^\infty~\prod_{b=1}^r\frac{(r-b)!\,
(1/2\lambda)^{4m_b}}{m_b!~\Gamma(m_b+n_b-b+r+1)}\nn\\&&
\times\,\prod_{a<b}\left(1+\frac{m_a-m_b+n_a-n_b}{b-a}\right) \ .
\label{calZdetexpl}\eea
Finally, we can simplify this expansion for ${\cal Z}_r$ even further
by decoupling the sum over partitions $\vn=(n_1,\dots,n_r)$. For this,
we define a new set of integers by
\bea
p_a&=&n_a-n_{a+1}+1 \ , ~~ a=1,\dots,r-1 \ , \nn\\p_r&=&n_r \ .
\label{pkdef}\eea
Then the $p_a$'s are all independent variables, constrained only by their
ranges which are given by $1\leq p_a<\infty$ for $a=1,\dots,r-1$ and
$-\infty<p_r<\infty$.

The decoupled expansion of the partition function is
thereby obtained by substituting
\beq
n_a=p_a+p_{a+1}+\dots+p_r+a-r \ ,
\label{nkpksub}\eeq
along with the explicit group theoretical formulas (\ref{dimirrep}) and
(\ref{C1}), into (\ref{calZdetexpl}) to get the final result (up to
irrelevant numerical factors)
\bea
{\cal Z}_r&=&\sum_{p_1=1}^\infty\cdots\sum_{p_{r-1}=1}^\infty~
\sum_{p_r=-\infty}^\infty\cos\left(\frac{2\pi\,q}L\,\sum_{b=1}^rb\,
p_b\right)\nn\\&&\times\,\sum_{m_1=0}^\infty\cdots
\sum_{m_r=0}^\infty~\prod_{b=1}^r\frac{(b-1)!\,
(2\lambda)^{-4m_b-2b\,p_b}}{m_b!~\Gamma(m_b+p_b+p_{b+1}+
\dots+p_r+1)}\nn\\&&\times\,\prod_{a<b}\,\frac{m_a-m_b+p_a+p_{a+1}+\dots+p_b}
{p_a+p_{a+1}+\dots+p_b} \ ,
\label{calZfinal}\eea
where we have used the reality of the left-hand side of (\ref{Wilsonchi}) to
make the expression for the partition function manifestly real by adding its
complex conjugate to itself. The partition function (\ref{calZfinal}) is a
straightforward expansion in powers of $1/\lambda^2$ over $2r$
independent integers $p_a,m_a$, $a=1,\dots,r$. Note the reduction in the
number of dynamical degrees of freedom of the model. The original
$2r^2$ degrees of freedom of the two-dimensional lattice gauge theory
(\ref{calZdef}) (or equivalently of the unitary two-matrix model
(\ref{calZTEK})) is reduced to $2r$. This proves that the lattice
model is exactly solvable, and thereby gives yet another indication
that noncommutative gauge theory in two dimensions is a topological
field theory. The sum (\ref{calZfinal}) is formally analogous to the
partition expansion of continuum noncommutative Yang-Mills theory.

\subsection{Scaling Limits}
\label{subsec:5.5}

The final step of this calculation should be to take the continuum
limit $\varepsilon\to0$ of the lattice theory. In order to prevent the
spacetime from degenerating to zero area, from (\ref{extent}) we see
that we must also take $L\to\infty$, or equivalently $r\to\infty$ in
(\ref{calZfinal}). There are different ways of performing these two
limits, each of which leads to a different continuum gauge theory. If
the limit is taken such that the dimensionful noncommutativity
parameter (\ref{thetalat}) vanishes, then the continuum limit is
ordinary Yang-Mills theory in two dimensions. The area (\ref{extent})
may be either finite or infinite in this limit. If $A\to\infty$, then
the expansion (\ref{calZpartitions}) truncates to the trivial
representation for which $n_a=0~~\forall a$ and one obtains
\beq
{\cal Z}_r\biggm|_{\stackrel{\scriptstyle\Theta\to0}{\scriptstyle
A\to\infty}}=\det_{a,b}\,\big[I_{b-a}(2/\lambda^2)\big] \ .
\label{calZrplane}\eeq
This expression is recovered in the naive
large $r$ limit due to the suppression of higher representations which
is induced by the dimension factors $d_{\vn}$ in the denominators of
(\ref{calZpartitions}). It is just the standard expression for
Yang-Mills theory on the plane which arises from the one-plaquette
Wilson model in the limit of a large number of
colours~\cite{BarsGreen}. Going back to (\ref{calZsumint}), we see
that the truncation to $\vn={\mbf 0}$ is indeed nothing but the Gross-Witten
reduction of commutative lattice gauge theory in two
dimensions~\cite{GrossWitten}.

The other scaling limit that one can take is $\varepsilon\to0$,
$L\to\infty$ with $\varepsilon^2L$ finite. Then the noncommutativity
parameter (\ref{thetalat}) is finite, but the area (\ref{extent})
diverges. The resulting continuum limit is gauge theory on the
noncommutative plane, and from (\ref{calZsumint}) we see that its
partition function generalizes that of ordinary Yang-Mills theory by
including a sum over non-trivial representations of the unitary
group. This quantitative difference is similar in spirit to that which
occurs in the group theory presentation of noncommutative gauge
theory~\cite{inprep2}, which can be thought of as a modification of ordinary
gauge theory by the addition of infinitely many higher Casimir operators to
the action (equivalently higher powers of the field strength
$F_A$). The inclusion of higher representations in the statistical sum
means that this series cannot be expressed in terms of a unitary
one-matrix model. Determinants such as (\ref{calZrplane}) whose matrix
elements depend only on the difference between row and column labels
are called Toeplitz determinants and are known to be equivalent to the
evaluation of a related unitary one-matrix integral~\cite{Itzykson}. In the
present noncommutative case, the partition function is not given by a Toeplitz
determinant, although it is represented by the unitary
two-matrix model (\ref{calZTEK}) and depends only on the eigenvalues
of the matrix $W=V_1\,V_2\,V_1^\dag\,V_2^\dag$.

Unravelling the precise continuum limit of the expansion
(\ref{calZfinal}) is one of the important unsolved analytical problems
in the combinatorial approach to two-dimensional noncommutative
Yang-Mills theory. The noncommutativity parameter $\Theta$ enters
in the 't~Hooft coupling constant as $\lambda^2=\pi\,g^2\Theta$ and
implicitly in the factors of $L=r/N$ appearing in (\ref{calZfinal}). It is
necessary to identify whether the double-scaling limit required, over
and above the naive continuum limit, exists within this
framework. Both the naive and non-trivial double-scaling limits have
been observed numerically in the Eguchi-Kawai model~\cite{Nish}, and more
recent numerical investigations indicate that they exist also within the full
noncommutative field theory~\cite{Bietenholz:2002ch},\cite{BHN1}--\cite{CAm}.
The rigorous derivation of this limit is
described at the classical level in~\cite{lls1}. Amongst other things,
the solution to this system may help in unravelling the mysterious
properties of the gauge group of noncommutative gauge theory, which in the
present context is formally an $r\to\infty$ limit of $U(r)$, confirming
other independent
expectations~\cite{ncproc1,gn,inprep},\cite{nair}--\cite{LSZ}. It would also be
interesting to understand the
complete solution of the discrete theory whose continuum spacetime is
a torus, which is given by a more general construction~\cite{AMNSFinite} to
which the present analysis does not apply.

\subsection*{Acknowledgments}

R.J.S. would like to thank the organisors and participants of the
meetings for the many questions and comments which have helped to
improve the material presented here, and also for the very
pleasant scientific and social atmospheres. He would also like to thank the
School of Theoretical Physics of the Dublin Institute for Advanced Study for
its hospitality during the completion of the manuscript. The work of R.J.S. was
supported in part by an Advanced Fellowship from the Particle
Physics and Astronomy Research Council~(U.K.).

%
%

%

\begin{thebibliography}{99.}
%
%
%

\bibitem{ncproc1}
L.~D.~Paniak and R.~J.~Szabo,
hep-th/0302195.

\bibitem{poly}
A.~P.~Polychronakos,
Phys. Lett. B {\bf 495}, 407 (2000)
[hep-th/0007043].

\bibitem{bak}
D.~Bak,
Phys. Lett. B {\bf 495}, 251 (2000)
[hep-th/0008204].

\bibitem{gn}
D.~J.~Gross and N.~A.~Nekrasov,
JHEP {\bf 0103}, 044 (2001)
[hep-th/0010090].

\bibitem{blp}
D.~Bak, K.~Lee and J.-H.~Park,
Phys. Rev. D {\bf 63}, 125010 (2001)
[hep-th/0011099].

\bibitem{bnt}
A.~Bassetto, G.~Nardelli and A.~Torrielli,
Nucl. Phys. B {\bf 617}, 308 (2001)
[hep-th/0107147].

\bibitem{guralnik}
Z.~Guralnik,
JHEP {\bf 0206}, 010 (2002)
[hep-th/0109079].

\bibitem{gsv}
L.~Griguolo, D.~Seminara and P.~Valtancoli,
JHEP {\bf 0112}, 024 (2001)
[hep-th/0110293].

\bibitem{Bietenholz:2002ch}
W.~Bietenholz, F.~Hofheinz and J.~Nishimura,
JHEP {\bf 0209}, 009 (2002)
[hep-th/0203151].

\bibitem{inprep}
L.~D.~Paniak and R.~J.~Szabo,
hep-th/0203166.

\bibitem{bnt2}
A.~Bassetto, G.~Nardelli and A.~Torielli,
Phys. Rev. D {\bf 66}, 085012 (2002)
[hep-th/0205210].

\bibitem{BassVian}
A.~Bassetto and F.~Vian,
JHEP {\bf 0210}, 004 (2002)
[hep-th/0207222].

\bibitem{torr}
A.~Torrielli,
hep-th/0301091.

\bibitem{inprep2} L.~D.~Paniak and R.~J.~Szabo,
hep-th/0302162.

\bibitem{ks}
A.~Konechny and A.~Schwarz,
Phys. Rept. {\bf 360}, 353 (2002)
[hep-th/0012145 , hep-th/0107251].

\bibitem{dougnek}
M.~R.~Douglas and N.~A.~Nekrasov,
Rev. Mod. Phys. {\bf 73}, 977 (2002) [hep-th/0106048].

\bibitem{sz1}
R.~J.~Szabo,
Phys. Rept. {\bf 378}, 207 (2003) [hep-th/0109162].

\bibitem{cmr}
S.~Cordes, G.~Moore and S.~Ramgoolam,
Nucl. Phys. Proc. Suppl. {\bf 41}, 184 (1995) [hep-th/9411210].

\bibitem{YMintegrable} A. Gorsky and N. A. Nekrasov, Nucl. Phys. B {\bf 414}, 213 (1994) [hep-th/9304047].

\bibitem{MinPol} J. A. Minahan and A. P. Polychronakos, Phys. Lett. B {\bf 312}, 155 (1993) [hep-th/9303153].

\bibitem{GPSS} G. Grignani, L. D. Paniak, G. W. Semenoff and P. Sodano, Ann. Phys. {\bf 260}, 275 (1997) [hep-th/9705102].

\bibitem{Wittensemi} E. Witten, J. Geom. Phys. {\bf 9}, 303 (1992) [hep-th/9204083].

\bibitem{Migdal} A. A. Migdal, Sov. Phys. JETP {\bf 42}, 743 (1975).

\bibitem{Rusakov} B. E. Rusakov, Mod. Phys. Lett. A {\bf 5}, 693 (1990).

\bibitem{amns}
J.~Ambj\o rn, Y.~M.~Makeenko, J.~Nishimura and R.~J.~Szabo,
JHEP {\bf 0005}, 023 (2000)
[hep-th/0004147].

\bibitem{WittenQ} E. Witten, Commun. Math. Phys. {\bf 141}, 153 (1991).

\bibitem{SzBook} R. J. Szabo, {\it Equivariant Cohomology and Localization of Path Integrals}, Lect. Notes Phys. {\bf m63} (Springer-Verlag, 2000).

\bibitem{AB} M. F. Atiyah and R. Bott, Phil. Trans. Roy. Soc. London A {\bf 308}, 523 (1982).

\bibitem{SW} N. Seiberg and E. Witten, JHEP {\bf 9909}, 032 (1999) [hep-th/9908142].

\bibitem{A-GB} L. Alvarez-Gaum\'e and J. L. F. Barb\`on, Nucl. Phys. B {\bf 623}, 165 (2002) [hep-th/0109176].

\bibitem{lls1}
G.~Landi, F.~Lizzi and R.~J.~Szabo,
Commun. Math. Phys. {\bf 217}, 181 (2001)
[hep-th/9912130].

\bibitem{CDS} A. Connes, M. R. Douglas and A. Schwarz, JHEP {\bf 9802}, 003 (1998) [hep-th/9711162].

\bibitem{SchwarzMorita} A. Schwarz, Nucl. Phys. B 534, 720 (1998) [hep-th/9805034].

\bibitem{Wilson} K. G. Wilson, Phys. Rev. D {\bf 10}, 2445 (1974).

\bibitem{GrossWitten} D. J. Gross and E. Witten, Phys. Rev. D {\bf 21}, 446 (1980).

\bibitem{AMNSFinite} J. Ambj\o rn, Y. M. Makeenko, J. Nishimura and R. J. Szabo, JHEP {\bf 9911}, 029 (1999) [hep-th/9911041].

\bibitem{EK} T. Eguchi and H. Kawai, Phys. Rev. Lett. {\bf 48}, 1063 (1982).

\bibitem{EKtwist} A. Gonzalez-Arroyo and C. P. Korthals Altes, Phys. Lett. B {\bf 131}, 396 (1983).

\bibitem{ID} J.-M. Drouffe and J.-B. Zuber, Phys. Rept. {\bf 102}, 1 (1983).

\bibitem{BarsGreen}
I.~Bars and F.~Green,
Phys. Rev. D {\bf 20}, 3311 (1979).

\bibitem{Itzykson} C. Itzykson, in: {\it Analytic Methods in Mathematical Physics} (Gordon and Breach, 1970), p. 469.

\bibitem{Nish} T. Nakajima and J. Nishimura, Nucl. Phys. B {\bf 528}, 355 (1998) [hep-th/9802082].

\bibitem{BHN1} W. Bietenholz, F. Hofheinz and J. Nishimura, hep-lat/0209021.

\bibitem{BHN2} W. Bietenholz, F. Hofheinz and J. Nishimura, hep-th/0212258.

\bibitem{CAm} J. Ambj\o rn and S. Catterall, Phys. Lett. B {\bf 549}, 253 (2002) [hep-lat/0209106].

\bibitem{nair}
V.~P.~Nair and A.~P.~Polychronakos,
Phys. Rev. Lett. {\bf 87}, 030403 (2001)
[hep-th/0102181].

\bibitem{harvey}
J.~A.~Harvey,
hep-th/0105242.

\bibitem{LSZ} F. Lizzi, R. J. Szabo and A. Zampini, JHEP {\bf 0108}, 032 (2001) [hep-th/0107115].

\end{thebibliography}
%



\end{document}